\documentclass[useAMS,usenatbib]{mn2e}
\usepackage{epsfig}
\usepackage[usenames,dvips]{color}
\usepackage{amsmath}
\usepackage{amssymb}

\newcommand{\Msolar}{\mbox{\,$\rm M_{\odot}$}} 
\newcommand{\Rsolar}{\mbox{\,$\rm R_{\odot}$}} 
\newcommand{\Lsolar}{\mbox{\,$\rm L_{\odot}$}} 
\newcommand{\Teff}{\mbox{\,\em T$_{\rm eff}$}} 
\def\simge{\mathrel{\raise1.16pt\hbox{$>$}\kern-7.0pt \lower3.06pt\hbox{{$\scriptstyle\sim$}}}} 
\def\simle{\mathrel{\raise1.16pt\hbox{$<$}\kern-7.0pt \lower3.06pt\hbox{{$\scriptstyle \sim$}}}} 

\newcommand{\iso}[2]{\mbox{$^{#1}{\rm #2}$}} 

\begin{document}

\title[CO+He WD mergers and R\,CrB evolution]{Post-merger evolution of carbon-oxygen + helium white dwarf binaries and the origin of R\,Coronae Borealis and extreme helium stars}
\author[]{Xianfei Zhang$^{1}$\thanks{E-mail: xiz@arm.ac.uk},  C. Simon Jeffery$^{1,2}$\thanks{E-mail: csj@arm.ac.uk}, Xuefei Chen$^{3,4}$, and Zhanwen Han$^{3,4}$\\
$^1$Armagh Observatory, College Hill, Armagh BT61 9DG, UK\\
$^2$School of Physics, Trinity College Dublin, Dublin 2, Ireland\\
$^3$Yunnan Observatories, Chinese Academy of Sciences, Kunming 650011, China \\
$^4$Key Laboratory for the Structure and Evolution of Celestial Objects, Chinese Academy of Sciences, Kunming, 650011, PR China}

\date{Accepted . Received ; in original form }

\pagerange{\pageref{firstpage}--\pageref{lastpage}} \pubyear{2011}

\maketitle

\label{firstpage}

\begin{abstract} Orbital decay by gravitational-wave radiation will cause some close-binary
white dwarfs (WDs) to merge within a Hubble time. The results from previous
hydrodynamical WD-merger simulations have been used to guide calculations of the
post-merger evolution of carbon-oxygen + helium (CO+He) WD binaries.
Our models include the formation of a hot corona in addition to a Keplerian disk.
We introduce a ``destroyed-disk'' model to simulate the effect of direct disk ingestion into the expanding envelope.
These calculations indicate significant lifetimes in the domain of the rare R Coronae Borealis (RCB) stars, before
a fast evolution through the domain of the hotter extreme helium (EHe)
stars. Surface chemistries of the resulting giants are in partial
agreement with the observed abundances of RCB and EHe stars.
The production of \iso{3}{He}, \iso{18}{O} and \iso{19}{F} are discussed.
Evolutionary timescales combined with binary white-dwarf merger rates from
binary-star population synthesis are consistent with present-day numbers of RCBs and EHes, provided
that the majority come from relatively recent ($<2$Gyr) star formation.
However, most RCBs should be produced by CO-WD + {\it low-mass} He-WD mergers, with the He-WD
having a mass in the range $0.20 - 0.35\Msolar$.  Whilst, previously, a high He-WD mass ($\geq 0.40 \Msolar$) was
required to match the carbon-rich abundances of RCB stars, the ``destroyed-disk" model yields a high-carbon product with
 He-WD mass  $\geq 0.30\Msolar$, in better agreement with population synthesis results.
\end{abstract}


\section{Introduction}
R Coronae Borealis (R\,CrB or RCB) stars are carbon-rich, hydrogen-poor supergiants \citep{Clayton96}. They
range in effective temperature ($T_{\rm eff}$) from 4\,000 to 8\,000 K. The surface gravities range
from $\log g = 0.5$ to 1.5. Due to the  ejection and condensation of carbon-rich dust clouds
(soot), RCBs show large visible brightness variations. The RCB population in the
Large Magellanic Cloud (LMC) provides estimates of absolute
magnitudes which range from Mv $\sim$2.5 to $\sim$5 and corresponding luminosities $\log
(L/\Lsolar)$ of 4.0 to 3.2 \citep{Alcock01}.
At present,  about 68 RCBs are known in the
Galaxy, 25 in the Magellanic Clouds, and 4 in M31
\citep{Clayton96, Zaniewski05, Tisserand08, Clayton2012, Tang2013}.

The surface compositions of RCB stars are extremely helium-rich (mass fraction 0.98). Besides
extreme hydrogen deficiency, RCB stars are enriched in N, Al, Na, Si, S, Ni, several
s-process elements, and sometimes O \citep{Lambert86, Asplund00}. The isotopic carbon ratio,
$^{12}\rm C/^{13}\rm C $, is very large ($>500$) in most RCB stars \citep{Clayton07}. In most
RCB stars where oxygen isotopic ratios can be measured, $^{18}\rm O $ is enhanced relative to
$^{16}\rm O $ by factors of 100$\sim$1000 compared to standard Galactic values. In addition, some
RCB stars are enhanced in lithium \citep{Asplund00}.

Extreme Helium stars (EHe) are early-type  supergiants with spectral types A and B
and atmospheres almost void of hydrogen, but
highly enriched in carbon \citep{Jeffery11}.
In many respects, their surface composition resembles that of the RCB stars,
although isotopic ratios and the abundances of some interesting  elements cannot be measured.
Because of these similarities,  a strong link between the cool RCB and the hot EHe stars has been postulated.

By considering these characteristics, two principle evolution channels have been established. A small
fraction may be produced following a late thermal pulse in a post-asymptotic giant-branch star on the
white dwarf cooling sequence \citep{Iben84a, Clayton11}. The majority are more likely to have been
produced following the merger of a carbon-oxygen white dwarf with a helium white dwarf
\citep{Webbink84,saio02,Jeffery11,Longland11,Staff2012, Menon2013,Dan2014}.
Less than 1\% of RCB stars may come from a high-mass double helium white dwarf merger \citep{Zhang12b}.

Using a simple nuclear reaction model, \citet{Clayton07} indicated that in a CO+He WD
merger, it is partial \iso{14}{N}+\iso{4}{He} burning at just the right temperature and duration that is the
likely generator of excess \iso{18}{O} (or a small \iso{16}{O}/\iso{18}{O} ratio).
In addition, the high values of $^{12}\rm C/^{13}\rm C $ observed
in RCB stars are probably produced by $3\alpha$ burning at the same time.
Using complete nucleosynthesis models for asymptotic giant
branch (AGB) stars as indicative of the composition of the CO white dwarf progenitor,
\citet{Jeffery11} examined the elements produced during a CO+He WD  merger;  the
results supported such a merger as a  possible progenitor of
RCB stars. Smoothed-particle hydrodynamics (SPH) simulations have sought
to probe the physics of the merger itself \citep{Yoon07,Loren09,Longland12,Zhu2013}.
Those SPH simulations indicated that before the slow or `cool' disk-accretion  phase \citep{saio02},
there would have been a  fast or `hot'  accretion  phase.
This fast phase may last between a few minutes
(in unsynchronized systems) and a few hours \citep{Raskin12,Dan2011},
during which approximately
half the accreted mass would form a hot corona around the more massive star. Since none
of the SPH calculations address the long-term post-merger evolution, and the \citet{saio02}
models did not incorporate a hot coronal phase, the picture of the post-merger
evolution of a CO+He WD binary remains incomplete.

In this paper, we construct one-dimensional stellar evolution models for a number of {\it dynamical}
CO+He WD mergers guided by the results of recent SPH calculations.  These models are evolved
through the RCB domain. The surface chemistries produced by the mergers and their subsequent
evolution are compared with observations of RCB stars and related objects.

\section{Carbon-oxygen + helium white dwarf mergers}

There are two key requirements for a close binary white dwarf to merge and form a single star.
The first is gravitational radiation whereby a close binary star system loses orbital angular
 momentum  $J_{\rm orb}$ and causes the binary separation $a$ to decay until the larger star fills
its Roche lobe. The rate of loss of orbital angular momentum \citep{Landau62} is expressed as

\[\begin{split}
\frac{\dot{J}_{\rm orb}}{J_{\rm orb}} & =
-8.3\times 10^{-10} \times  \\
 & \left( \frac{m_1}{\Msolar}\right) \left(\frac{m_2}{\Msolar}\right)
\left(\frac{m_1+m_2}{\Msolar}\right)\left(\frac{a}{\Rsolar}\right)^{-4}{\rm yr}^{-1},
\end{split}\]
where $m_1$ and $m_2$ are the masses of stars in the binary.

\begin{figure}
\centering \includegraphics [angle=0,width=0.45\textwidth]{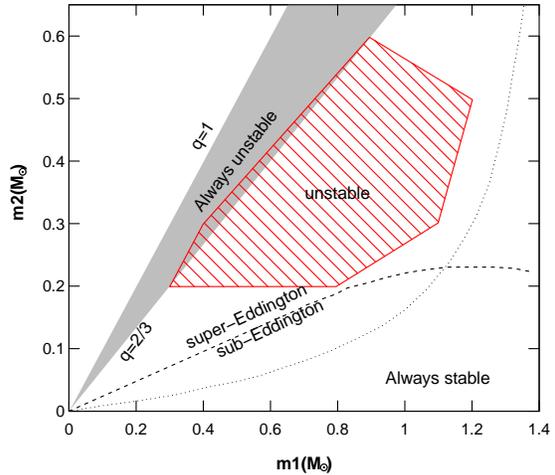}
\caption{Mass transfer in double WD binaries. Binaries below the dotted line will transfer mass by stable Roche-Lobe overflow (RLOF)
and evolve into stable disk-accreting AM CVn binaries. The dashed line separates sub- and
super-Eddington accretion systems. Mass transfer in the grey region will be  unstable and
will lead to dynamical mergers. The area between the solid ($q=2/3$) and dashed lines corresponds to
either stable or unstable mass transfer depending on the spin-orbit coupling. The hatched region marks the area
studied by \citet{Dan2011}.}
\label{binary}
\end{figure}

\begin{figure*}
\centering
\includegraphics [angle=0,scale=0.5]{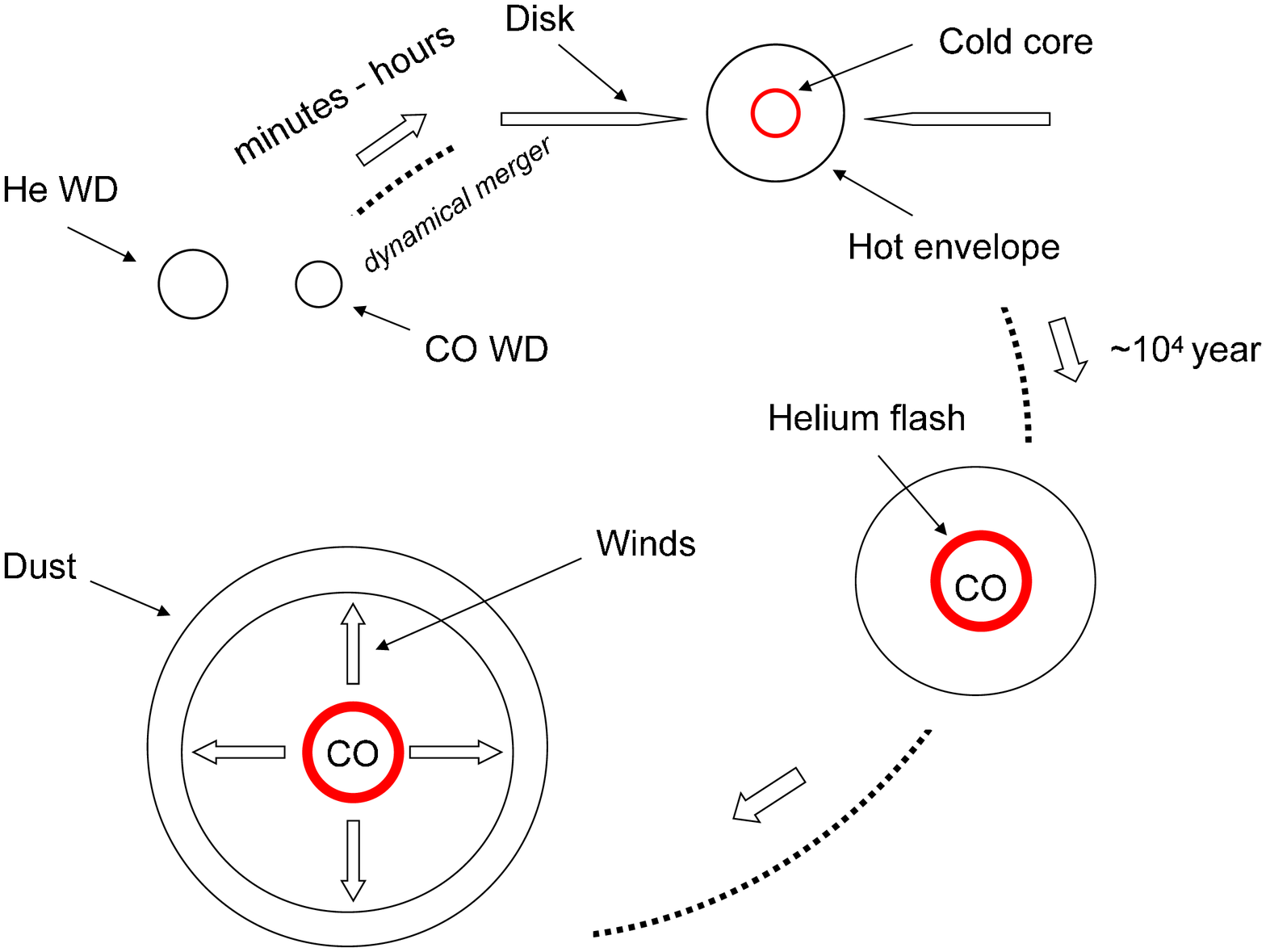}
\caption{Schematic of possible steps in a carbon-oxygen + helium white dwarf merger leading to the formation an RCB star. }
\label{merger}
\end{figure*}

The second requirement is the mass ratio $q$. If, at the point where the larger (less massive)
white dwarf fills its Roche  lobe and starts to lose mass,
   \[
    q \equiv m_2/m_1 \geq q_{\rm crit}  \equiv \frac{5}{6}+ \frac{\zeta(m_2)}{2},
    \]
where $\zeta(m_2) \equiv {\rm d} \ln r/ {\rm d} \ln m$ is obtained from the white-dwarf mass-radius
relation, its radius will increase more quickly than the
separation will increase due to the transfer of angular momentum,
i.e. the synchronisation timescale of spin-orbit coupling, $\tau_s \rightarrow 0$. This leads to unstable (runaway) mass
transfer on a dynamical timescale (a few seconds), corresponding to the grey zone in Fig.~\ref{binary}.
However, \citet{Nelemans2001} point out that there is a much more stringent condition under which
no angular momentum is transferred from accretor to donor,
i.e. the synchronisation timescale of spin-orbit coupling, $\tau_s \rightarrow \infty$.
Under this condition,  $q_{\rm crit}$ could decrease to 1/5, as shown by the dotted line in Fig.~\ref{binary}.
If $q \leq q_{\rm crit}$, stable mass transfer
will occur, possibly leading to the formation of an AM\,CVn binary - an ultrashort binary system
containing two helium white dwarfs. Thus, white dwarf  binaries with mass ratios in the range $2/3 < q <  1$ will merge dynamically.
In the region below the dotted line, mass transfer is always stable.
The region between both zones corresponds to either stable or unstable mass transfer depending on the spin-orbit
coupling \citep{Marsh2004}. The dashed line in Fig.~\ref{binary} separates sub- and super-Eddington accretion.
Under strong coupling, numerical simulations \citep{Motl07} indicate that $ q<2/3$
results in stable mass transfer, $2/3<q<0.9$ leads to merger and tidal disruption of the donor, and for
$q>0.9$, the donor plunges into the accretor. Under weak coupling, the simulation of \citet{Dan2011} shows that
some low mass-ratio binaries will be unstable and experience super-Eddington accretion (hatched zone of Fig.~\ref{binary}).

Hydrodynamical simulations of the merger of two low-mass white dwarfs suggest a number of phases
occur. Following the complete tidal disruption of the lower mass component (secondary, $m_2$), its
material may be redistributed around the more massive component (primary, $m_1$) in (a) a cold
Keplerian debris disk and (b) a hot spherical corona. A disk allows cold mass to migrate towards
its centre, from where it can be accreted slowly onto the primary surface, while angular momentum is
dissipated towards the disk circumference \citep{Lynden74}. Fig.~\ref{merger} shows a simplified
sequence of events during a CO + He WD merger as suggested by hydrodynamical simulations. First, the
He white dwarf is disrupted by the CO white dwarf in a few minutes, and forms a cold Keplerian
debris disk and a hot spherical corona. Then, mass accretion starts from the disk to the star; this
process may take around $10^{4}$ years. Once the accreted envelope is massive enough, stable
helium-shell burning is ignited and the star becomes a cool giant. At large radius and low \Teff, a
stellar wind, pulsation and soot formation will give the star the characteristics of an R\,CrB variable.
In addition, a number of RCB stars show nebular emission and extended low-temperature
dust shells \citep{Clayton96}. These are not seen in EHe stars. It is not yet clear whether this
circumstellar material is related to the merger process or not.

\begin{table}
\caption{Summary of the `corona+disk'  accretion models described in \S 3.2, showing for each experiment
the mass of the carbon-oxygen white dwarf  $m_{\rm CO}$,
the helium white dwarf $m_{\rm He}$, the corona mass $m_{\rm corona}$, the disk mass $m_{\rm disk}$, and the final mass of the merger product $m_{\rm final}$, all in solar  units.} 
\begin{tabular}{cccccc}
\hline
 Model            & $m_{\rm CO}$ & $m_{\rm He}$ & $m_{\rm corona}$  & $m_{\rm disk}$ & $m_{\rm final}$ \\
 \hline
 1                  & 0.50 & 0.40 & 0.23 & 0.17 & 0.90\\
 2                  & 0.50 & 0.45 & 0.31 & 0.14 & 0.95\\
 3                  & 0.55 & 0.35 & 0.15 & 0.20 & 0.90\\
 4                  & 0.55 & 0.40 & 0.20 & 0.20 & 0.95\\
 5                  & 0.55 & 0.45 & 0.25 & 0.20 & 1.00\\
 6                  & 0.60 & 0.30 & 0.10 & 0.20 & 0.90\\
 7                  & 0.60 & 0.35 & 0.12 & 0.23 & 0.95\\
 8                  & 0.60 & 0.40 & 0.16 & 0.24 & 1.00\\
 \hline
 \end{tabular}
 \label{table:models} 
\end{table}

\section{Models}
\subsection{Numerical simulations}

Numerical simulations of stellar evolution have been
carried out using the stellar evolution code ``Modules for Experiments in Stellar
Astrophysics'' (MESA) version  4028 \citep{paxton11}.
Carbon-oxygen white-dwarf models were generated by
evolving a 3$\Msolar$ main-sequence star and removing the envelope when this reached the
required mass (see \citet{Zhang12a} for details). The post-merger evolution was modeled
by considering fast accretion at $10^4 \rm \Msolar yr^{-1}$, representing the formation
of a hot corona, followed by slow accretion at $10^{-5} \rm \Msolar yr^{-1}$,
representing accretion from a Keplerian disk. After the accretion process, we employ
a Bl\"ocker-type stellar wind of the form:
\[\begin{split}
P_0 & < 100 {\rm d}: \\
\dot{M} & = \dot{M}_{\rm R} \equiv 4\times10^{-13} \eta_{\rm R} \frac{LR}{M}  \frac{\Msolar}{\Lsolar\Rsolar}  {\rm \Msolar yr^{-1}  }
\end{split}\]
and
\[\begin{split}
P_0 & > 100 {\rm d}: \\
\dot{M} & = 4.83\times10^{-9} \left(\frac{M_{\rm ZAMS}}{\Msolar}\right)^{-2.1}
\left(\frac{L}{\Lsolar}\right)^{2.7} \dot{M}_{\rm R} {\rm  \Msolar yr^{-1}  }
\end{split}\]
where the fundamental-mode pulsation period $P_0$ is approximated by
\[
P_0 = 0.012 \left(\frac{R}{\Rsolar}\right)^{1.86} \left(\frac{M}{\Msolar}\right)^{-0.73} {\rm  d}
\]
\citep{Bloecker1995}, and where we adopt two alternative values for the Reimers' coefficient $\eta_{\rm R} = 0.02$ and  0.1.
 The value normally adopted for Galactic giants is 0.1. Adopting the smaller value of 0.02 makes some sense for C-rich giants,
but the effect of hydrogen deficiency on the mass-loss efficiency is not at all known.
The value 0.02 is in agreement with the luminosity and mass loss rates of some lithium and
carbon-rich AGB stars in the Magellanic Clouds given by \citet{Ventura2000,Ventura2013},
and used for the modelling the thermally-pulsing AGB phase \citep{Lau2008}.
The composition of the accreted material is assumed to be $Y=0.98, Z=0.02$, where the detailed metal
abundance distribution is obtained from the average abundance of the He WD model. We set the ratio
of mixing length to local pressure scale height, $\alpha = l/H_{\rm p} = 2.0$. The opacity tables
for various metallicities are those compiled by \citet{Grevesse1998} and \citet{Ferguson2005}.
The default model atmosphere is an Eddington grey photosphere.
The model resolution is determined by an adaptive grid in mass and an adaptive timestep.
 Nuclear burning is computed and applied explicitly for
every timestep in the evolution calculation. Rotation is not treated in the version of MESA used. 
The evolution calculations do not yet follow the entire evolution through the EHe phase owing to
to numerical instabilities associated with the very narrow helium-burning shell.

\subsection{Hot corona and disk masses: $m_{\rm corona}, m_{\rm disk}$}

As \citet{Zhang12b, Zhang12a} discussed, the final surface composition of the
merger depends strongly on the relative mass fractions in the hot corona and the disk.
\citet{Zhu2013} studied a detailed parameter space for
double carbon-oxygen white dwarf mergers with an SPH simulation,
restricted to the systems without tidal locking.
They give a formula for the masses of the accretor ($m_1$) and
core-corona ($m_{\rm ce}=m_1+m_{\rm corona}$) as:
\[
       \frac{m_{\rm ce}}{m_1}=1+0.81q_\rho   \pm 0.03 ,
\]
 where $q_\rho$ is the central density ratio of the donor ($m_{2}=m_{\rm disk}+m_{\rm corona}$) and accretor ($m_1$), $q_\rho \equiv \rho_{c,2}/\rho_{c,1}$.
 In the mass range of 0.4 to 1.0  \Msolar, the central density $\rho_c$ depends approximately exponentially on mass,
\[
       \rho_c =3.3\times10^7 e^{5.64\times(m -1)} \rm {g\,cm^{-3}},
\]
 where $m$ is in solar masses.
 As very little mass is ejected during the simulation (the merger is essentially conservative),
the coronal mass is:
 \[
      m_{\rm corona} =0.81\times e^{5.64\times(m_2 -m_1)} \times m_1 \pm 0.03 \times m_1,
 \]
 and
 \[
      m_{\rm disk} =m_2- m_{\rm corona}.
 \]

Although our calculations are for a CO+He WD merger, the masses are within the range of
the \citet{Zhu2013} parameter space.  We have therefore used this result to provide the
coronal and disk masses in the eight models calculated for this paper and which are shown in Table ~\ref{table:models}.

Based on the corona-plus-disk model,
a $\it ``destroyed-disk" model$ will be introduced in \S7,
in which the disk is assumed to be totally destroyed
on a convective turnover timescale.

\begin{figure}
\centering \includegraphics [angle=0,width=0.45\textwidth]{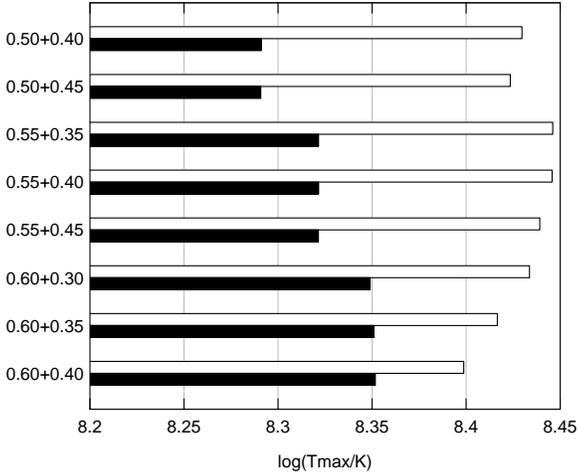}
\caption{The maximum temperatures achieved in  each of the `corona+disk' merger models.
The empty bar indicates the maximum temperature after the fast accretion phase.
The filled bar indicates the maximum temperature after the slow accretion phase.
The vertical axis labels identify each model by CO+He WD mass. }
\label{tp}
\end{figure}

\begin{figure}
\centering \includegraphics [angle=0,width=0.45\textwidth]{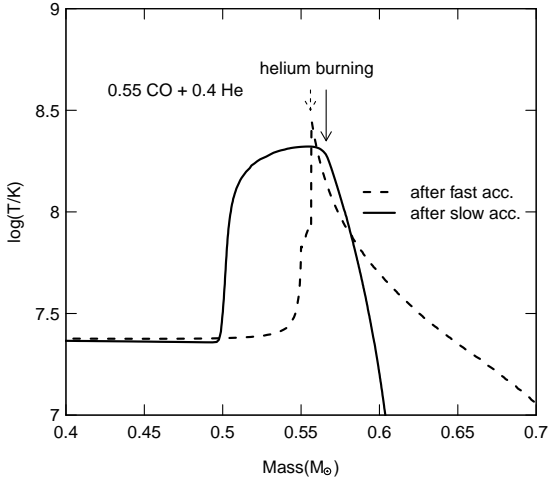}
\caption{The temperature profile of model 0.55+0.40 \Msolar\ immediately after the  fast accretion phase (dashed line)
and  after the slow accretion phase (solid). The arrows indicate the location of the center of $3\alpha$ burning.}
\label{055tp}
\end{figure}

\begin{figure}
\centering \includegraphics [angle=0,width=0.45\textwidth]{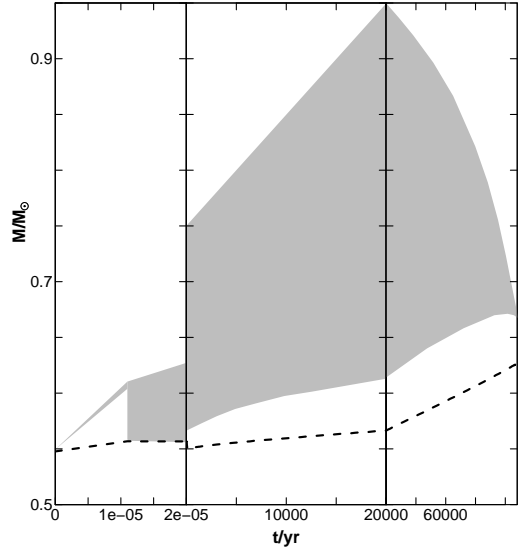}
\caption{Convection regions during evolution (grey zone). Left: the location of convective regions
during fast accretion of helium from a 0.40 \Msolar\ helium white dwarf onto 0.55 $\Msolar$
carbon-oxygen white dwarf. The fast accretion rate is $10^{4} \Msolar {\rm yr}^{-1}$. Middle: the
location of convective regions during the slow accretion of helium to the final mass 0.95 \Msolar.
The slow accretion rate is $10^{-5} \Msolar {\rm yr}^{-1}$. Right: the location of convective
regions during the post-merger evolution of the 0.95 \Msolar\ product with mass loss via a
Reimers'-type stellar wind ($\eta_{\rm R} = 0.02$). The dashed line shows the helium-burning
location.}
\label{convection}
\end{figure}

\begin{figure}
\centering \includegraphics [angle=0,width=0.45\textwidth]{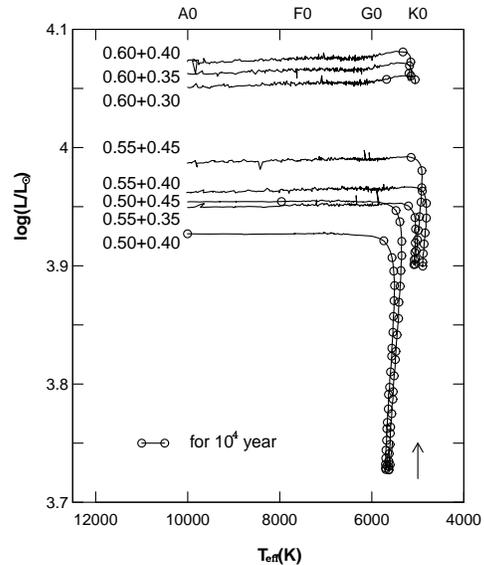}
\caption{The $L-\Teff$ diagram showing post-merger evolution of all `corona+disk' models. Circles represent intervals of $10^4$ yrs.
The arrow indicates the evolutionary direction, which is towards higher luminosity. }
\label{multrcb}
\end{figure}

\section{Evolution tracks}

In the SPH simulations of white dwarf mergers, matter in the hot corona may, for some models,
briefly reach temperatures of $6\rm\times10^8 K$ or more \citep{Yoon07,Loren09}, so that some
nucleosynthesis of $\alpha$-rich material will occur. For models comparable to those calculated
here, \citet{Zhu2013} find peak temperatures $\approx 1.74-2.47 \rm\times 10^8$K. Our
one-dimensional quasi-equilibrium calculations indicate coronal temperatures up to $2.5\rm\times10^8
K$ for all models after the fast accretion phase (see details in Fig.~\ref{tp}). At these
temperatures, the 3$\alpha$ and other alpha-capture reactions are ignited at the surface of the
accretor almost immediately. Thus, carbon is produced by helium burning and nitrogen is destroyed by
$\alpha$ capture. As \citet{Warner67} and \citet{Clayton07} indicated, the destruction of
\iso{14}{N} by the $\iso{14}{N}(\alpha,\gamma)\iso{18}{O}$ reaction becomes more efficient than it
is at low temperature. As the temperature continues to increase, \iso{18}{O} produces \iso{22}{Ne}
through an additional $\alpha$ capture. Because of the high accretion rate, there is no runaway
helium flash and burning is stable. Most of the energy released during this process goes into
heating the corona.

After the fast-accretion phase, slow accretion transfers matter from the disk to the accretor,
whilst inward heat conduction from the helium-burning  shell continues to heat the CO core.
Meanwhile, the star starts to expand to become a red giant. After the slow accretion phase, the
temperature maximum decreases, and the width of the hot shell increases. An example of the
temperature profile of model 4 (0.55+0.40 \Msolar) at two different epochs is shown in
Fig.~\ref{055tp}.

The location of convective regions during the accretion and post-merger phases of model 4
(0.55+0.40 $\Msolar$) is shown in Fig.~\ref{convection}. After the fast accretion phase, the
increasing nuclear luminosity forces the star to expand; the radius becomes about 210 \Rsolar\
within 500 yrs. At this point, the envelope is almost fully convective, and helium-burning makes the
corona rich in $\iso{12}{C}$, $\iso{18}{O}$ and $\iso{22}{Ne}$. The surface abundance of $\iso{12}{C}$
comes from the 3$\alpha$ reaction, which is very sensitive to temperature. However, during the (slow) disk-accretion
phase the new ash of $\iso{12}{C}$ cannot reach the convection region, as shown in
Fig.~\ref{convection}. Thus, all carbon enrichment occurs during the fast-accretion phase.

After the slow-accretion phase, the star becomes more luminous, and the mass-loss rate starts to
become significant. The mass-loss rate depends on the Reimers' coefficient, and will be discussed in
the next section.

As the helium shell burns outwards through the accreted envelope, and the wind removes mass from the
surface, the helium envelope is eventually not massive enough to sustain a helium-burning shell, and
the shell luminosity starts to drop. To compensate, the envelope shrinks and the star evolves to
become a white dwarf, evolving to higher \Teff\ at constant luminosity. The evolution tracks for
each model in the $L - \Teff$ diagram are shown in Fig.~\ref{multrcb}.

The evolution calculations were terminated when the contracting star reached $\Teff=10\,000\,{\rm K}$.
This limit is currently imposed by numerical instabilities associated with the extremely narrow helium-burning
shell and the algorithm used by MESA to select an appropriate timestep. Work to overcome this
limitation is in progress.


\begin{figure}
\centering \includegraphics [angle=0,width=0.45\textwidth]{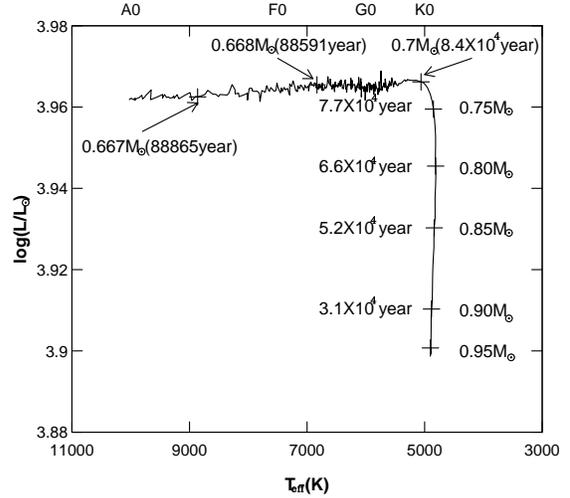}
\caption{Detail of the $L-\Teff$ diagram for the 0.55+0.40\Msolar\ post-merger model ($\eta_{\rm R} = 0.02$). Times since the
end of the slow-accretion phase and the total mass at those times are shown at key points on the track.}
\label{055hr}
\end{figure}

\begin{figure}
\centering \includegraphics [angle=0,width=0.45\textwidth]{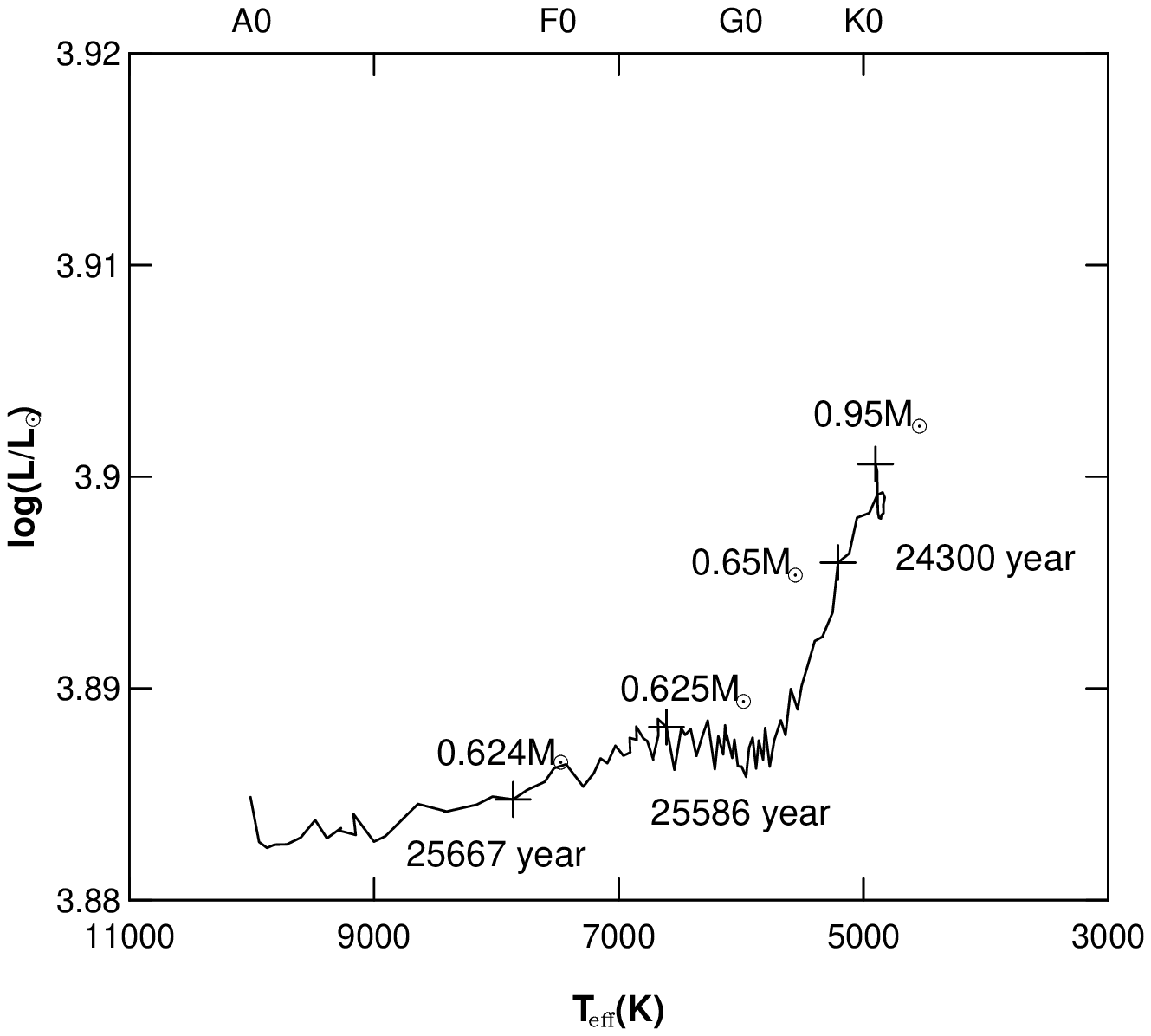}
\caption{Detail of the $L-\Teff$ diagram for the 0.55+0.40\Msolar\ post-merger model ($\eta_{\rm R} = 0.1$). Times since the
end of the slow-accretion phase and the total mass at those times are shown at key points on the track.}
\label{055hr01}
\end{figure}

\begin{figure}
\centering \includegraphics [angle=0,width=0.45\textwidth]{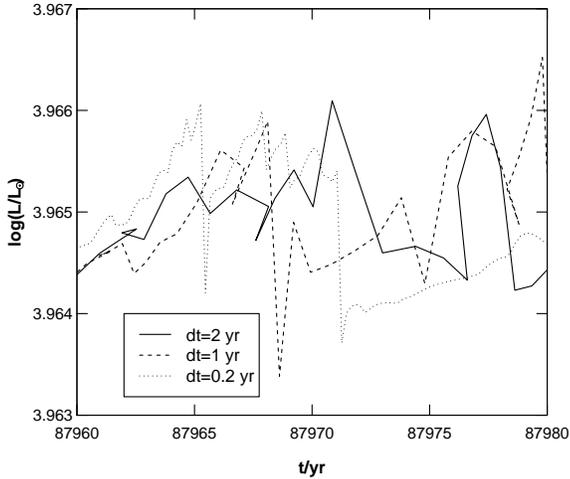}
\caption{Detail of the luminosity changes for the 0.55+0.40\Msolar\ post-merger model ($\eta_{\rm R} = 0.02$). The solid, dashed and dotted lines for evolutionary tracks of timesteps of 2, 1 and 0.2 years respectively.}
\label{noise}
\end{figure}

\section{Mass loss and dust formation}

A carbon-rich dust environment is an important feature of RCB variables,
the formation of which leads to the large visible brightness variations which are characteristic of the class.
In our simulations, we assumed a Bl\"ocker-type stellar wind \citep{Bloecker1995}, of the type usually
used  to represent mass loss in AGB stars. For the longest-lived phase of our models, this gave a  mass-loss rate
$\sim 10^{-6}\Msolar {\rm yr^{-1}}$, which is comparable with observational estimates \citep{Clayton2012}.
But this mass-loss rate is essentially determined by the Reimers' coefficient  which
has been set to  $\eta_{\rm R}=0.02$ (see above). For $\eta_{\rm R}=0.1$, our model gave a  mass-loss rate
$\sim 10^{-5}\Msolar {\rm yr^{-1}}$, and luminosity more than $\sim 1500 \Lsolar$ lower than the $\eta_{\rm R}=0.02$ model.
{
Details from the $L-\Teff$ diagram for the 0.55+0.40$\Msolar$ post-merger model are
shown in Figs.~\ref{055hr} and ~\ref{055hr01}, for $\eta_{\rm R}=0.02$ and  0.1 respectively.
The $\eta_{\rm R}=0.02$ model loses 0.25\Msolar\  within $8.4\times10^4$ yrs,
and the $\eta_{\rm R}=0.1$ model loses 0.3\Msolar\  within $2.5\times10^4$ yrs.
Both models show small random changes of luminosity during mass loss.  We repeated the
evolution calculation for  20 years  of the $\eta_{\rm R}=0.02$ model using different timesteps (Fig.~\ref{noise}).
The small-amplitude variations did not repeat themselves identically  (although still present),
so we conclude that this `noise' is principally numerical.

In view of the observational constraints on mass-loss rate and luminosity provided by the RCB stars, we chose
$\eta_{\rm R}=0.02$ for more detailed investigation.
The evolution can be divided into two stages.
During the first  stage (A),
the luminosity increases as the CO-core mass increases, with virtually no change in \Teff.
In the following 5000 yrs (stage B), where the  star contracts at constant luminosity (see above),
 the mass loss rate reduces to $6\times10^{-7}\Msolar {\rm yr^{-1}}$, and the star loses a further 0.03\Msolar\
to give a final mass $\approx0.67\Msolar$. We note that the mass is less than the estimate of
$0.8-0.9 \Msolar$ from pulsation modeling \citep{saio02,Saio2008}, but that the mass-loss rate is
higher than that estimated for EHe stars by \citet{Jeffery10}.
If we assume a gas-to-dust ratio of 100 \citep{Whittet2003,Clayton2012},
we estimate that  $\approx 2.5\times10^{-4}\Msolar$ of  dust would be produced during
stage A ($8.4\times10^4\,{\rm yr}$).


\begin{figure}
\centering \includegraphics [angle=0,width=0.45\textwidth]{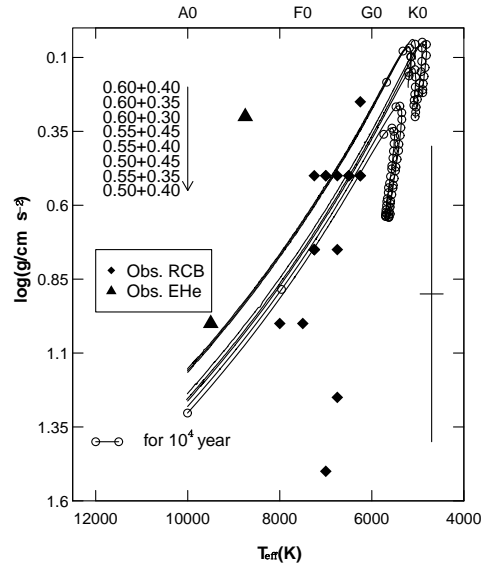}
\caption{The surface-gravity -- effective-temperature diagram for post-merger models. The diamonds show observed RCB stars.
The triangles show  observed EHe stars. The right cross shows an average error bar for the observations.}
\label{multgt}
\end{figure}

\begin{figure}
\centering \includegraphics [angle=0,width=0.45\textwidth]{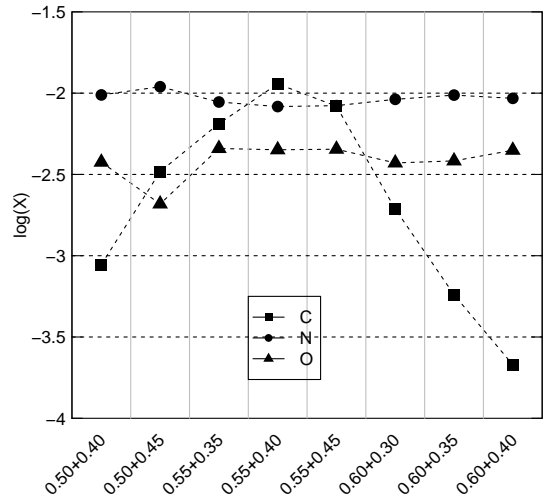}
\caption{The surface carbon (squares), nitrogen  (circles) and oxygen (triangles) abundances of each model.  }
\label{cno}
\end{figure}

\begin{table*}
\caption[Post-merger surface abundances]{Logarithmic mass fraction of carbon ($\beta_{\rm C}$),
nitrogen ($\beta_{\rm N}$), oxygen ($\beta _{\rm O}$), fluorine ($\beta_{\rm F}$),
neon ($\beta_{\rm Ne}$), lithium ($\beta_{\rm Li}$) and the number ratios for $\iso{12}{C}/\iso{13}{C}$, and $\iso{16}{O}/\iso{18}{O}$.
Three models with accreted-hydrogen mass fractions $X=0$ and $X=0.0001$
are shown.   Solar abundances are from \citet{AG89}. }
\centering
\begin{tabular}{lcccccccc}
\hline
  Model &$\beta_{\rm C}$&$\beta_{\rm N}$&$\beta_{\rm O}$&$\beta_{\rm F}$&$\beta_{\rm Ne}$&$\beta_{\rm Li}$&$\iso{12}{C}/\iso{13}{C}$&$\iso{16}{O}/\iso{18}{O}$\\ [0.5ex]	
\hline
\hline
  $X=0$\\
\hline
   0.55+0.35 \Msolar&-2.18&-2.05&-2.34&-7.47&-2.60&-7.25&284&0.56\\
   0.55+0.40 \Msolar&-1.94&-2.08&-2.34&-7.26&-2.45&-7.11&497&0.60\\
   0.55+0.45 \Msolar&-2.08&-2.07&-2.34&-7.21&-2.48&-7.05&360&0.57\\
\hline
  $ X=0.0001$\\
\hline
   0.55+0.35 \Msolar&-3.04&-2.01&-2.46&-4.81&-2.66&-7.73&28&0.85\\
   0.55+0.40 \Msolar&-3.08&-1.94&-2.80&-4.81&-2,73&-7.81&29&224\\
   0.55+0.45 \Msolar&-2.30&-2.00&-2.77&-3.94&-2.39&-7.79&136&12\\
\hline
   Solar&-2.51&-2.96&-2.02&-6.39&-2.76&-8.02&90&497\\
\hline
 \end{tabular}
 \label{table:modelscp} 
\end{table*}

\section{Surface Composition}

Fig.~\ref{multgt} shows that most
of the RCB stars lie around the part of the post-merger evolution
corresponding to the  final contraction phase (stage B).
To compare with our simulations,
data on the chemical composition of 17 RCB and 2  EHe stars with $T_{\rm eff} < 10\,000 {\rm K}$
have been taken from the compilation by \citet{Jeffery11}, which,  in turn,
was based on measurements by \citet{Asplund00} and \citet{Pandey08}.
Data for  $\Teff$ and surface gravities ($\log g$) of the same stars were
taken from the latter two sources.

\begin{figure*}
\centering \includegraphics [angle=0,width=0.6\textwidth]{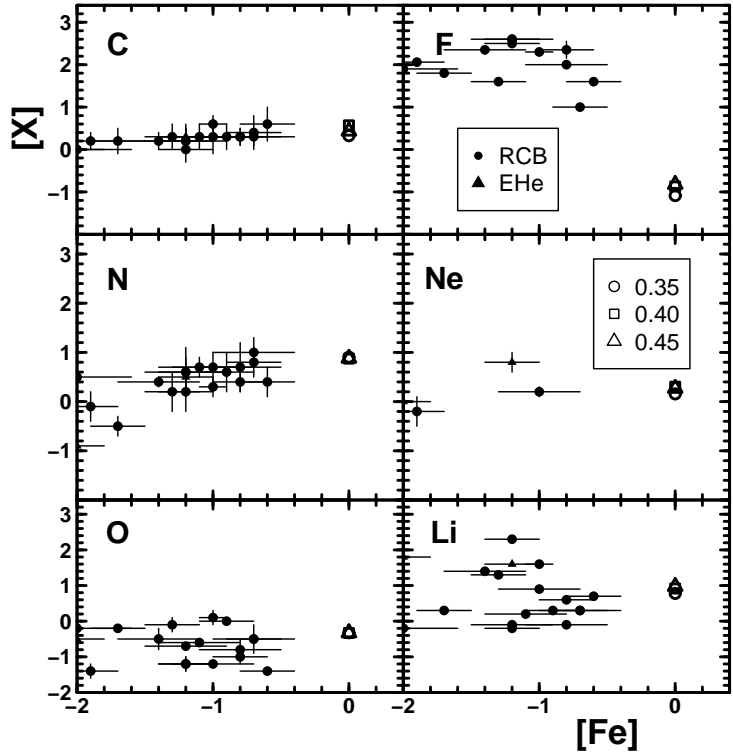}
\caption{The observed surface abundances of RCB and EHe stars \citep{Jeffery11,Asplund00,Pandey08}
compared with our nucleosynthesis computation (with accreted $X=0$). The axes [X] and [Fe] give logarithmic
abundances {\it by number}  relative to solar \citep{AG89} for individual elements and for iron, respectively.
Open circles, squares and diamonds are abundances given by our simulations for
0.55+0.35 \Msolar, 0.55+0.40 \Msolar\ and 0.55+0.45 \Msolar respectively.}
\label{abundance}
\end{figure*}

\begin{figure}
\centering \includegraphics [angle=0,width=0.45\textwidth]{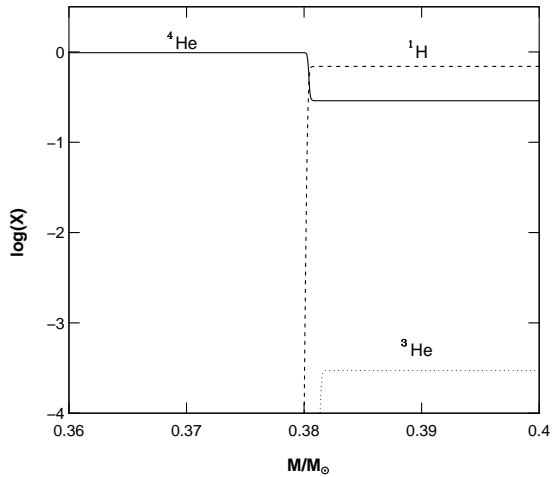}
\caption{The abundance profile of a 0.4 \Msolar\ He white dwarf with a 0.02 \Msolar\ hydrogen envelope.
The solid, dashed and dotted lines show \iso{4}{He}, \iso{1}{H} and \iso{3}{He}, respectively.}
\label{heprofile}
\end{figure}

\begin{figure}
\centering \includegraphics [angle=0,width=0.45\textwidth]{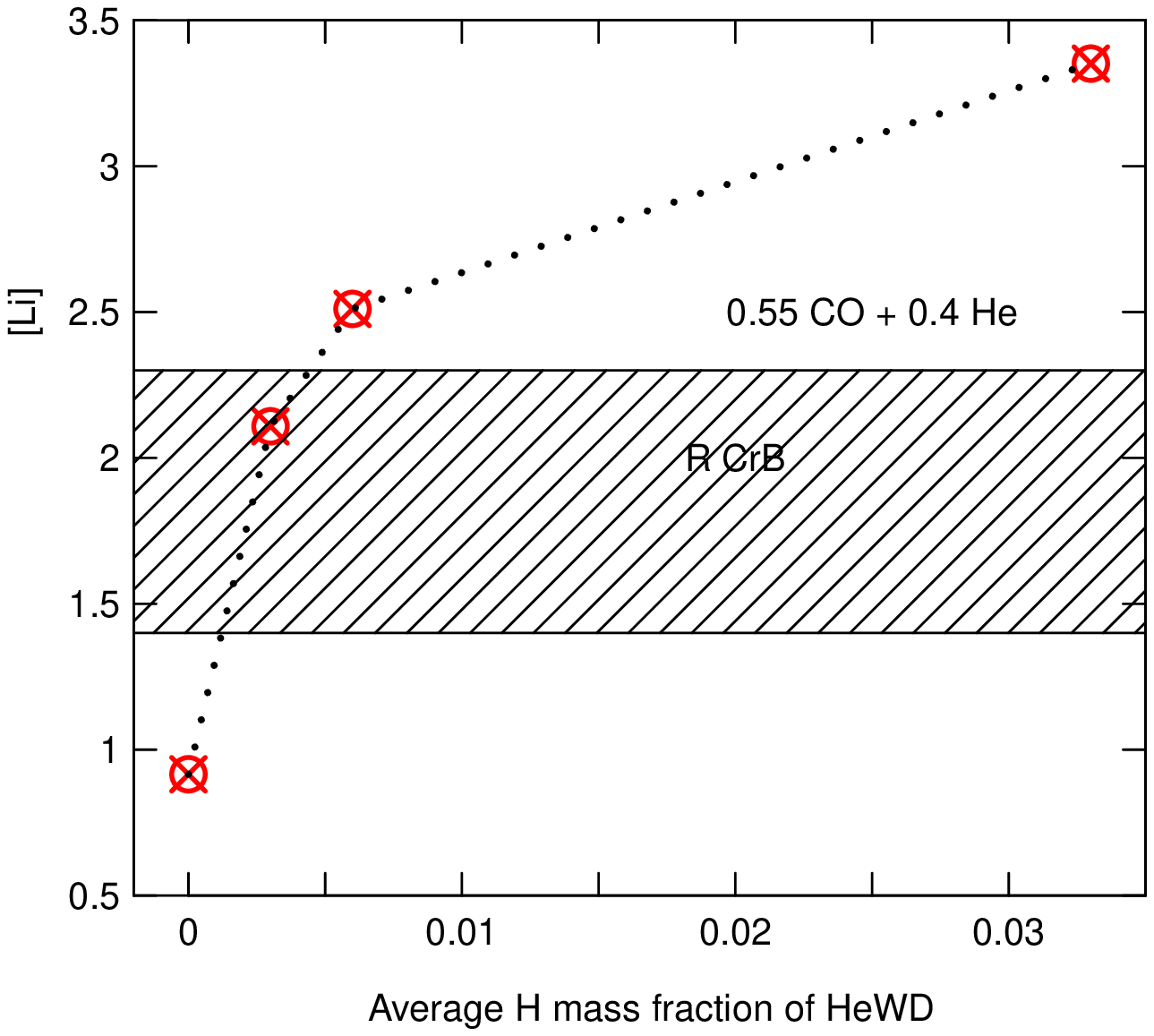}
\caption{The abundance of Li produced during merger as a function of the hydrogen mass fraction in the accreted He white dwarf.
The shaded zone shows the range of observed RCB lithium abundances.}
\label{lium}
\end{figure}

\subsection{Individual elements}

At the end of fast accretion and the beginning of slow accretion, flash-driven convection
mixes \iso{12}{C} throughout the envelope. During the subsequent slow-accretion phase,  coronally-produced
material is buried by C-poor material from the disk, so the new surface abundance very much depends on
whereabouts in the star the disk material is deposited.
In our simulations, which deposit disk material onto the surface of the accretor,
only the models for a 0.55$\Msolar$ CO white dwarf merger with a helium white dwarf
can produce a carbon-rich surface; see Fig.~\ref{cno} for details.
All of the $\iso{12}{C}$ enrichment comes from the fast accretion phase as discussed above.
The final carbon abundance depends on the peak temperature during accretion; in the models studied here,
mergers onto a 0.55$\Msolar$ CO white dwarf show the highest surface carbon. This appears
counter-intuitive. However, because of the larger mass ratio, the 0.60\Msolar\ CO white dwarf accretes less mass
during the fast merger phase, and hence the corona is cooler than in the 0.55\Msolar\ case (Fig.\,\ref{tp}), leading to the
production of {\it less} carbon. 

Since we are concerned with the origin of the carbon-rich RCB stars, we
focus our discussion on the three models producing most carbon, {\it i.e.} 0.55+0.35 \Msolar, 0.55+0.40 \Msolar\ and 0.55+0.45 \Msolar.
The principal surface abundances arising from these models are shown in Fig.~\ref{abundance} and Table ~\ref{table:modelscp}.

\paragraph*{Carbon}  is enriched in all RCB stars.
In those cool enough to show CO, the \iso{12}{C}/\iso{13}{C} ratios are very large ($\approx500$),
indicating a $3\alpha$ or helium-burning origin for the carbon excess. In our models,
\iso{12}{C} was produced by helium burning through the $3\alpha$ reaction during the hot
accretion phase, and then brought up to the surface during the slow-accretion phase
by flash-driven convection. The \iso{13}{C} abundance remained almost unchanged during the whole
simulation. This enrichment of \iso{12}{C} results in a very high ratio of
\iso{12}{C}/\iso{13}{C}, being 284, 497, and 360 in our simulation
for models for  0.55+0.35 \Msolar, 0.55+0.40 \Msolar\ and 0.55+0.45 \Msolar, respectively.
The total carbon, including \iso{12}{C} and \iso{13}{C},
shows an abundance similar to the observation (Fig.~\ref{abundance}).

\paragraph*{Nitrogen} is enriched in the RCB
stars. \citet{heber83} and subsequent authors point out that the N abundances in general
follow the trend of the iron abundance. Nitrogen is enriched through CNO cycling in the parent stars.
In our simulation, it is subsequently reduced by the $\iso{14}{N}(\alpha,\gamma)\iso{18}{O} $
reaction.

\paragraph*{Oxygen} is enriched in most RCB stars.
\citet{Clayton07} report that, for four RCBs, \iso{18}{O} has a large enrichment and shows a low
ratio of $\iso{16}{O}/\iso{18}{O}$ close to unity. By comparison with the solar abundance,
\iso{18}{O} must have increased by a factor $> 400$. \citet{Clayton07} also indicate that the production of
\iso{18}{O} requires temperatures of at least $10^8$ K to allow the
$\iso{14}{N}(\alpha,\gamma)\iso{18}{O} $ reaction. \iso{18}{O} is also produced by
$\alpha$-capture on \iso{14}{N} during the fast-accretion phase in our models;
the resulting $\iso{16}{O}/\iso{18}{O}$ ratios are 0.56, 0.60, and 0.57
for  models 0.55+0.35 \Msolar, 0.55+0.40 \Msolar\ and 0.55+0.45 \Msolar, respectively.
There is no {\it new} $\iso{16}{O}$ from alpha-capture during the fast-accretion phase.

\paragraph*{Fluorine} is enriched in most RCB stars by factors of 800--8000 relative to its probable initial
abundance \citep{Pandey08}. In our simulation, \iso{19}{F} comes from
 the reaction  $\iso{15}{N}(\alpha,\gamma)\iso{19}{F}$,
but the quantity of \iso{19}{F} produced is small.
The newborn \iso{19}{F} is mixed with disk-accreted material.
As \iso{19}{F} was destroyed by $\iso{19}{F}(p,\alpha)\iso{16}{O}$ during  RGB evolution,
the helium white dwarf is deficient in \iso{19}{F}, having a mass fraction
$\approx 8.5\times10^{-9}$ in our experiments. The fresh \iso{19}{F}
produced during the merger in our models is insufficient to match the observations.

\paragraph*{Neon:} a high overabundance of neon has been identified in several RCBs. In
our simulation, \iso{22}{Ne} is enriched by two $\alpha$-captures on \iso{14}{N} followed
by extensive convective mixing.

\paragraph*{Lithium:} a few RCB stars have a notably large overabundance of lithium,
which has so far been difficult to explain. \citet{Longland12} show that lithium can be produced by the
merging of a He WD with a CO WD if their chemical
composition is rich in \iso{3}{He} from the previous evolution. This
model requires enough \iso{3}{He} to be left in the white dwarf after the end of
main-sequence evolution and a hot enough corona to form during the merger.
In our simulation, we obtained a \iso{3}{He}
mass fraction of  $7.7\times10^{-8}$ in the He WD,
which subsequently yielded a post-merger surface with
a lithium mass fraction of $5.6\times10^{-8}, 7.7\times10^{-8}$, and $8.8\times10^{-8}$
for models 0.55+0.35 \Msolar, 0.55+0.40 \Msolar\ and 0.55+0.45 \Msolar, respectively,
or about ten times solar.
Hence it is clear that \iso{3}{He} is the key to explain the RCB lithium abundances.

\begin{figure*}
\centering \includegraphics [angle=0,scale=0.8]{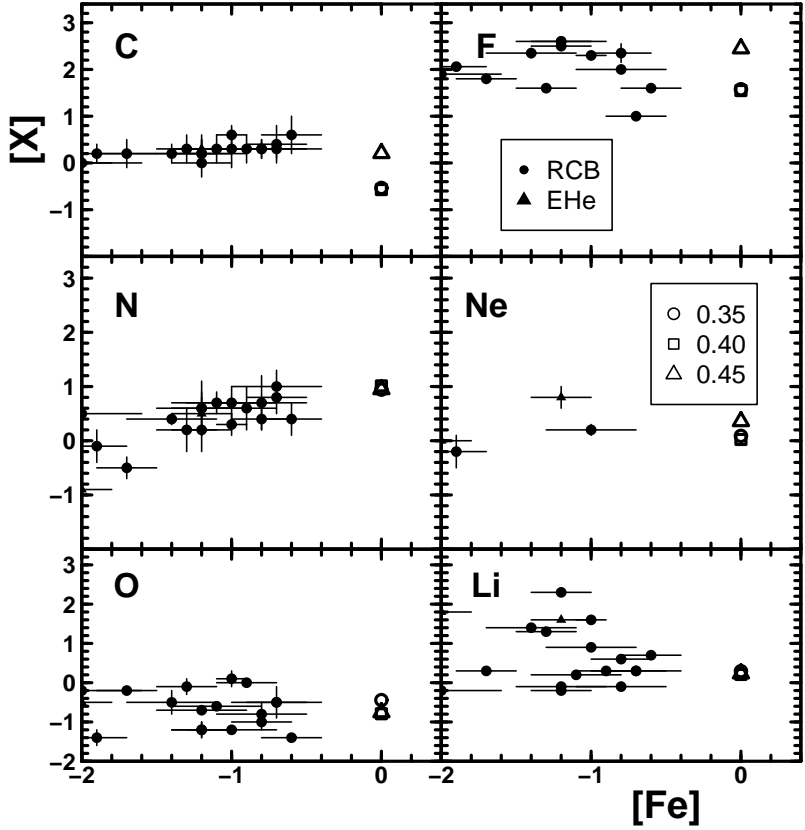}
\caption{As Fig.~\ref{abundance}, with hydrogen accretion ($X=0.0001$). The
observed surface abundances of RCB and EHe stars \citep{Jeffery11,Asplund00,Pandey08} compared with our
nucleosynthesis computation. The axes [X] and [Fe] give logarithmic
abundances  {\it by number} relative to solar for individual elements and for iron, respectively.
Open circles, squares and diamonds are abundances given by our simulations for
models 0.55+0.35 $\Msolar$, 0.55+0.40 $\Msolar$ and 0.55+0.45 $\Msolar$, respectively.}
\label{abundance2}
\end{figure*}

\subsection{The r\^ole of the helium white dwarf envelope}

According to stellar evolution theory, \iso{3}{He} in a helium white
dwarf is thought to be produced whilst the star was previously on the first giant branch
where it is dredged into and remains in the hydrogen envelope
 \citep{Benvenuto1998,Althaus2000,Althaus2001,Steinfadt2010,Longland12}.
Thus, the more of the hydrogen envelope that survives from the giant to the helium white dwarf,
the more \iso{3}{He} it is expected to contain. The He WD envelope plays
a very important r\^ole in nucleosynthesis during the merger, particularly with regard to
the production of lithium and fluorine.

To test whether there is  a relation between hydrogen abundance and lithium enrichment,
we set another three 0.4 \Msolar\ He WDs,
which have hydrogen envelope masses of 0.002, 0.004, 0.02\Msolar,
to merge with a 0.55 \Msolar\ CO WD.
The abundance profile of a 0.4 \Msolar\ He WD with a 0.02 \Msolar\ hydrogen envelope is shown in Fig.~\ref{heprofile}.
We assumed that all coronal hydrogen would be destroyed
 in the very initial stage of the fast-accretion phase,
and so set this phase to be hydrogen free ($X=0$), but maintained the contribution from \iso{3}{He}.
Without this assumption (i.e. with $X\ne0$), stable helium-burning could not be achieved
 during the accretion phase.
Hydrogen which is not destroyed immediately at the beginning of accretion
 will be completely mixed in the accretion materials and may survive in both the
disk and outer envelope.
 In a second experiment, we therefore tested the inclusion of a hydrogen mass
fraction $X=0.0001$ during both accretion phases.

 As shown in Fig.~\ref{abundance2} and Table ~\ref{table:modelscp},
the abundances for nitrogen, neon and oxygen are similar to models
without hydrogen.
The carbon abundance is lower in models including hydrogen than in models without hydrogen.
Carbon is decreased by the $\iso{12}{C}(p,\gamma)\iso{13}{N}(,\beta^+\mu)\iso{13}{C}(p,\gamma)\iso{14}{N}$
reaction in all models where hydrogen is present in the accreted material. At the same time,
the $\iso{12}{C}/\iso{13}{C}$ ratio is also decreased and does not agree with observation.
Hence the observed $\iso{12}{C}/\iso{13}{C}$ ratios limit the amount of hydrogen that can be present
in the accreted material.

The more massive the hydrogen envelope in the helium white dwarf,
 the more \iso{3}{He} it contains, and hence the more lithium is produced in
the merger (Fig.~\ref{lium}).

The inclusion of a small quantity of hydrogen also affects the
final fluorine abundances via the proton capture reaction \iso{18}{O}(p,$\gamma$)\iso{19}{F}.
Since a surplus of \iso{18}{O} is produced during the hot coronal phase, and only a small
quantity of \iso{19}{F} is required to substantially alter the initial abundances, this mechanism
 can explain the fluorine excesses observed in RCB and EHe stars \citep{Pandey06,Pandey08}.

The above tests demonstrate that a hydrogen envelope on the helium white dwarf
readily explains  the enrichment of both \iso{7}{Li} and \iso{19}{F}.
The abundance of \iso{7}{Li} depends on the fraction of \iso{3}{He} in the hydrogen envelope;
and the abundance of \iso{19}{F} depends on the fraction of \iso{1}{H} in the accreted material.

\begin{figure}
\centering \includegraphics [angle=0,width=0.3\textwidth]{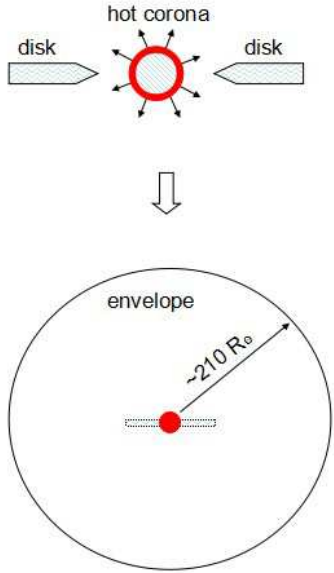}
\caption{Cartoon illustrating possible envelope expanding phase of merger.}
\label{expand}
\end{figure}

\begin{figure}
\centering \includegraphics [angle=0,width=0.45\textwidth]{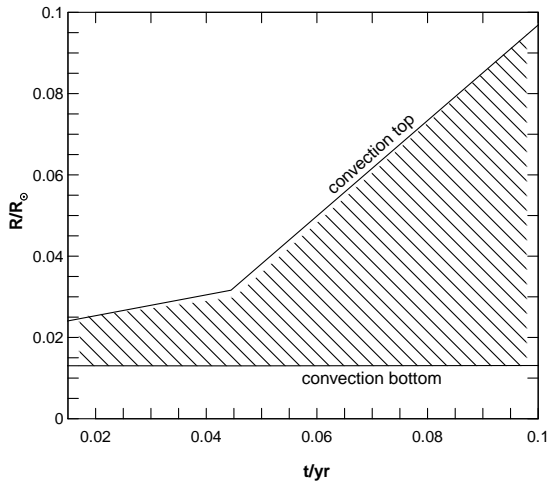}
\caption{ Radius evolution of the convection zone of a  0.55+0.40\Msolar destroyed-disk merger
model immediately following  the fast accretion phase. }
\label{rcov}
\end{figure}

\section{Disk Destruction}

If we only consider evolution after the fast-accretion phase, the star will  expand to become a giant on a
short timescale. For instance, the 0.55+0.20 \Msolar\ merger model expands to a radius of $\sim$ 210\Rsolar
within 500 years (Fig.~\ref{expand}).

SPH simulations show the the debris disks to be smaller
than 0.1\Rsolar\ \citep{Yoon07,Loren09,Longland12,Zhu2013,Dan2014}.
The envelope becomes almost fully convective when expanding, and hence the Keplerian disk may well be destroyed
by convection and other mixing processes. Driven by helium-burning at its base, the convection region of the envelope will expand to
0.1\Rsolar\ within 0.1 years (Fig.~\ref{rcov}).
If the disk is assumed to be totally destroyed by convection, the accretion
rate will be $2 \Msolar\,{\rm yr}^{-1}$.  We adopt this accretion rate to make a new class of post-merger model, which
we call ``destroyed disk'' models, and evolve these models with $\eta_{\rm R}=0.02$ stellar winds, and  hydrogen-free ($X=0$) accretion.

In order to consider both destroyed-disk and low-$q$ mergers, we produced 15 models comprising 0.50, 0.55, and 0.6 $\Msolar$ CO WDs
merging with 0.20, 0.25, 0.30, 0.35, and 0.40 $\Msolar$ He WDs. We set the mass-accretion rate to be  $2 \Msolar\,{\rm yr}^{-1}$ to emulate the
disk-destruction process.

\begin{figure}
\centering \includegraphics [angle=0,width=0.45\textwidth]{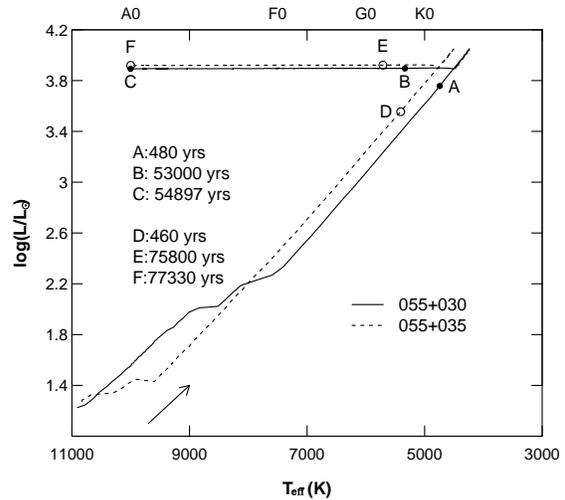}
\caption{Evolutionary tracks of two destroyed-disk models. The solid and dashed lines for
0.55+0.30 and 0.55+0.35 \Msolar\ models with $\eta_{\rm R}=0.02$, respectively. Letters A--C and D--F indicate times since merger for each of the two tracks. }
\label{055hrhotwarm}
\end{figure}

\begin{figure*}
\centering \includegraphics [angle=0,width=0.9\textwidth]{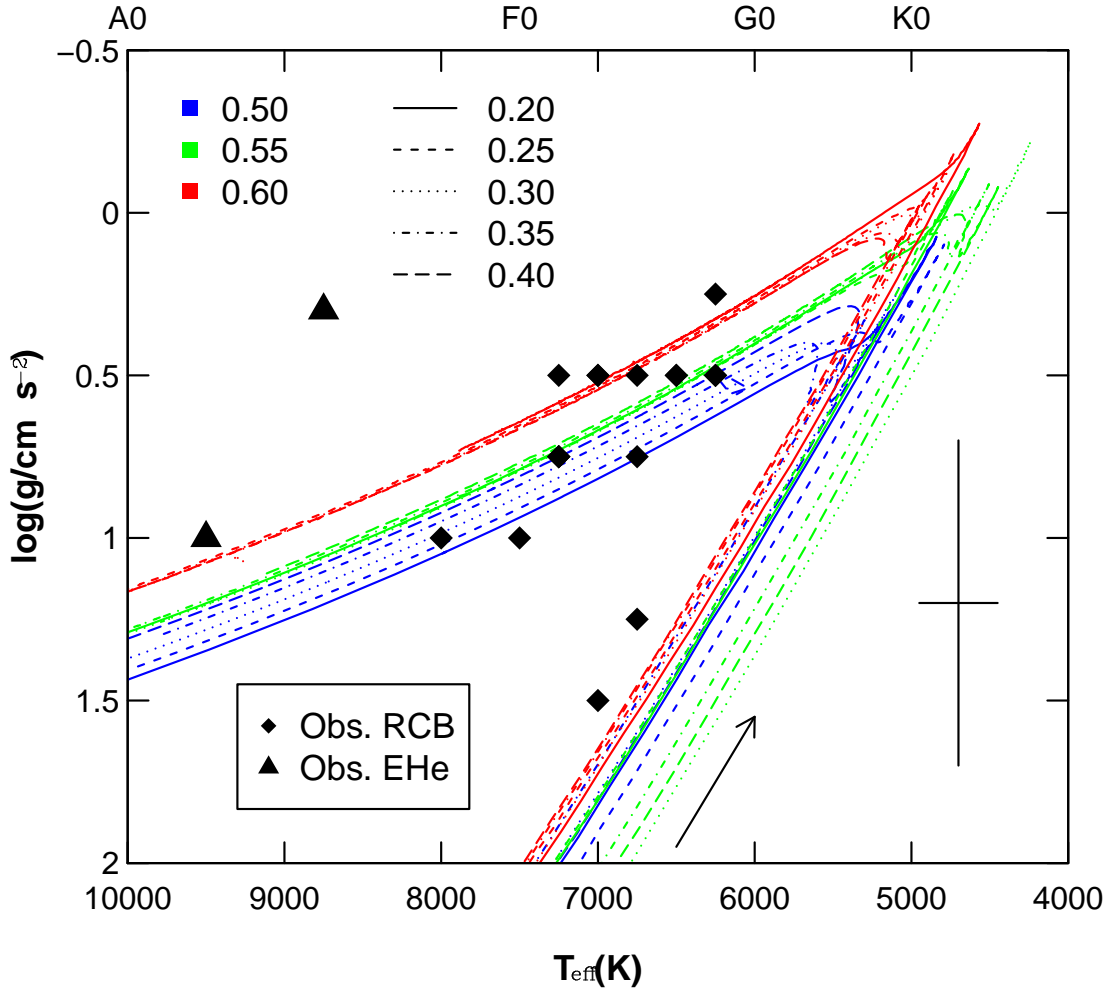}
\caption{The surface gravity -- effective temperature tracks for destroyed-disk  post-merger models.
Different colours identify different CO WD masses, {\it i.e.} blue for 0.50 $\Msolar$, green for 0.55 $\Msolar$
and red for 0.60 $\Msolar$. Different line styles identify different  He WD masses, {\it i.e.} solid for 0.20 $\Msolar$,
dashed for 0.25 $\Msolar$, dotted for 0.30 $\Msolar$, dash-dotted for 0.35 $\Msolar$,
and long-dashed for 0.40 $\Msolar$. The arrow indicates the evolutionary direction, which is initially towards lower gravity.
The diamonds show observed RCB stars. The triangles show observed EHe stars with $T_{\rm eff} < 10\,000 {\rm K}$.
The cross at the lower right shows the average error in the observations. }
\label{multgtlow1}
\end{figure*}

\begin{table*}
\caption[Post-merger surface abundances]{Post-merger surface abundances for disk-destroyed models with accreted $X=0$. The table shows
logarithmic mass fractions $\beta$  of carbon ($\beta_{\rm C}$),
nitrogen ($\beta_{\rm N}$), oxygen ($\beta _{\rm O}$), fluorine ($\beta_{\rm F}$),
neon ($\beta_{\rm Ne}$), lithium ($\beta_{\rm Li}$) and the number ratios for $\iso{12}{C}/\iso{13}{C}$, and $\iso{16}{O}/\iso{18}{O}$.
Solar abundances are from \citet{AG89}. }
\centering
\begin{tabular}{lcccccccc}
\hline
  Model (CO+He) &$\beta_{\rm C}$&$\beta_{\rm N}$&$\beta_{\rm O}$&$\beta_{\rm F}$&$\beta_{\rm Ne}$&$\beta_{\rm Li}$&$\iso{12}{C}/\iso{13}{C}$&$\iso{16}{O}/\iso{18}{O}$\\ [0.5ex]	
\hline
\hline
   0.50+0.20\Msolar&-2.23&-2.16&-2.20&-7.31&-2.49&-6.94&259&0.35\\
   0.50+0.25\Msolar&-2.09&-2.14&-2.20&-7.28&-2.53&-6.87&361&0.35\\
   0.50+0.30\Msolar&-2.50&-2.17&-2.13&-7.58&-2.65&-6.75&140&0.28\\
   0.50+0.35\Msolar&-3.24&-2.04&-2.32&-7.94&-2.79&-6.75&24&0.51\\
   0.50+0.40\Msolar&-2.32&-1.96&-2.68&-7.55&-2.65&-6.93&208&3.12\\
\hline
   0.55+0.20\Msolar&-2.27&-2.22&-2.19&-7.28&-2.34&-6.98&235&0.34\\
   0.55+0.25\Msolar&-2.46&-2.41&-1.99&-7.41&-2.47&-6.81&151&0.19\\
   0.55+0.30\Msolar&-1.69&-2.26&-2.26&-6.83&-2.17&-6.78&891&0.45\\
   0.55+0.35\Msolar&-1.89&-2.12&-2.25&-7.19&-2.47&-6.75&571&0.43\\
   0.55+0.40\Msolar&-1.78&-2.08&-2.34&-7.16&-2.46&-6.75&735&0.60\\
\hline
   0.60+0.20\Msolar&-2.42&-2.21&-2.16&-7.45&-2.42&-7.07&164&0.31\\
   0.60+0.25\Msolar&-3.04&-2.22&-2.09&-7.77&-2.57&-6.97&39&0.25\\
   0.60+0.30\Msolar&-2.81&-2.16&-2.15&-7.58&-2.62&-6.94&66&0.30\\
   0.60+0.35\Msolar&-3.55&-2.06&-2.27&-7.95&-2.78&-6.94&11&0.43\\
   0.60+0.40\Msolar&-3.75&-1.98&-2.50&-8.01&-2.82&-6.89&6&1.03\\
\hline
   Solar&-2.51&-2.96&-2.02&-6.39&-2.76&-8.02&90&497\\
\hline
 \end{tabular}
 \label{table:warm} 
\end{table*}

\subsection{Hertzsprung-Russell  diagram}
The post-merger evolutionary tracks of two  destroyed-disk models with different masses are shown in Fig.~\ref{055hrhotwarm}.
After the merger process, they take $\sim$ 500 years to expand to the giant position.
After the Reimers' wind has removed most of the mass from the envelope (typically $\approx 70$\%), shell-helium burning makes the star evolve to
high temperature without a luminosity change, as in most post-AGB stars.
Taking the 0.55+0.30 \Msolar\ model, for instance, the star takes 480 years to expand to 112 \Rsolar\ (marked A on Fig.~\ref{055hrhotwarm}) and
only loses $3.3\times 10^{-6}$ \Msolar. Evolution from A to B takes 52\,520 years, with a loss of  0.22 \Msolar, and the radius only decreases to 103 \Rsolar.
Evolution from stage B to C takes 1897 years and the temperature increases from 5336 K to 10000 K.

\begin{figure}
\centering \includegraphics [angle=0,width=0.45\textwidth]{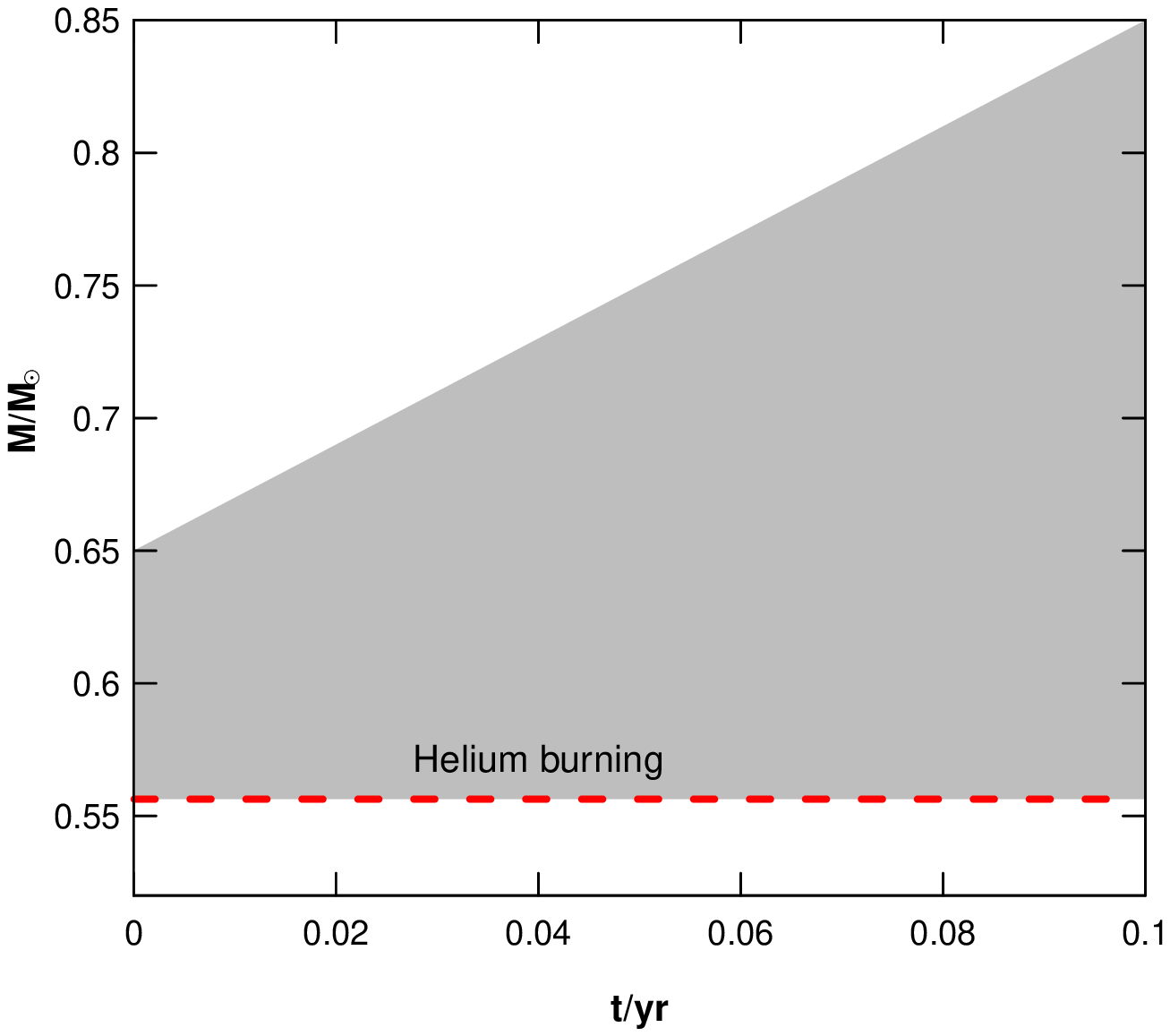}
\caption{Convection boundaries during the early evolution of the 0.55+0.30 \Msolar\ destroyed-disk model.}
\label{dd-conv}
\end{figure}

\subsection{Comparison with observation}
The locations of seventeen RCB and two EHe stars with $T_{\rm eff} < 10\,000 {\rm K}$
on the temperature-gravity diagram are compared with the destroyed-disk
post-merger evolution tracks (Fig.~\ref{multgtlow1}).
In this figure, most RCB stars lie around the part of the post-merger evolution
corresponding to the  final contraction phase (B to C).

\subsection{Abundances}
Table~\ref{table:warm}  shows that the surface carbon abundance varies with mass.
We also find enhanced abundances of  N, O, Ne, and Li.
All of these elements  are compared with observed RCB and EHe stars  in
Fig.~\ref{abundance6}. Fluorine is low because there was no hydrogen in the accreted material.
These calculations show that carbon is, again, only enhanced to the degree observed in models
where the CO WD has a mass $\approx 0.55\Msolar$. However, high surface carbon is achieved for
{\it lower} He WD masses, {\it i.e.} $\geq 0.30 \Msolar$, than in the models presented in \S 4--6.

The slow-neutron capture process (s-process) occurs at temperatures $T > 2.2-3.5 \times 10^8 {\rm K}$
and $\rho > 1-3\times 10^3 {\rm g/cm}^3$ \citep{Raiteri1991a,Meyer1994}.
The temperature and density of the burning shell in our models satisfies this condition. In the destroyed-disk model,
convection completely mixes the outer layers of the star so that any s-processed material produced in the shell
during the merger and/or disk ingestion will become visible on the surface (Fig.~\ref{dd-conv}).
However, s-process nucleosynthesis is not included in the version of MESA used in these calculations, so we
are not in a position to make a comparison with the observational constraints.

\begin{figure*}
\centering \includegraphics [angle=0,scale=0.8]{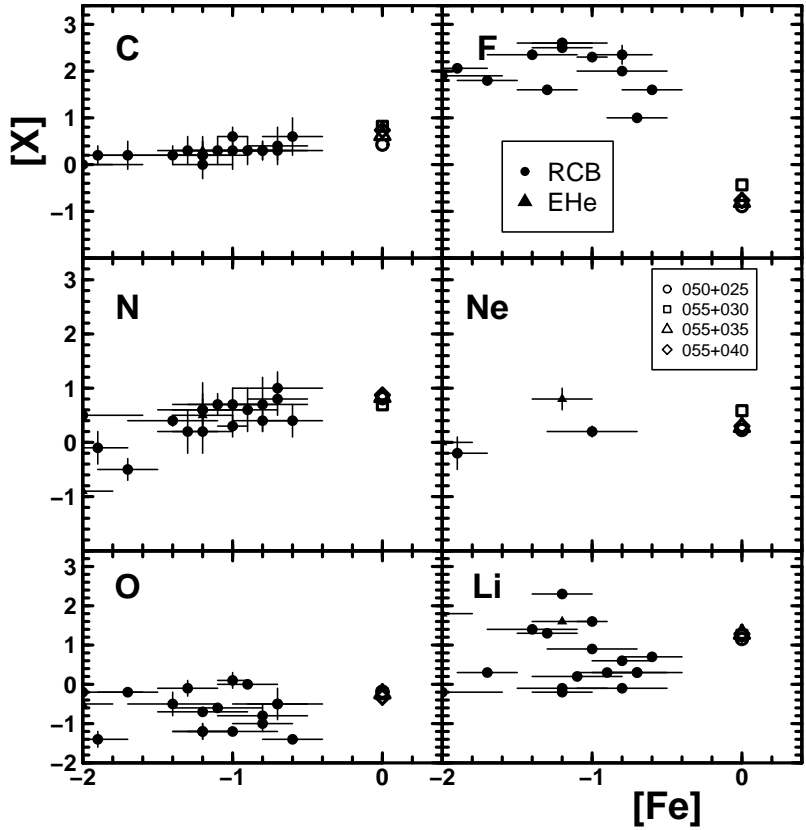}
\caption{The observed surface abundances of RCB and EHe stars \citep{Jeffery11,Asplund00,Pandey08}
compared with our nucleosynthesis computation for destroyed-disk models (accreted $X=0$). The axes [X] and [Fe] give logarithmic
abundances {\it by number}  relative to solar \citep{AG89} for individual elements and for iron, respectively.
Open circles, squares, triangle and diamonds are abundances for the 0.55+0.25, 0.55+0.30 \Msolar, 0.55+0.35 \Msolar\ and 0.55+0.40 \Msolar\
models, respectively.}
\label{abundance6}
\end{figure*}

\begin{figure}
\begin{center}
\includegraphics[width=0.45\textwidth]{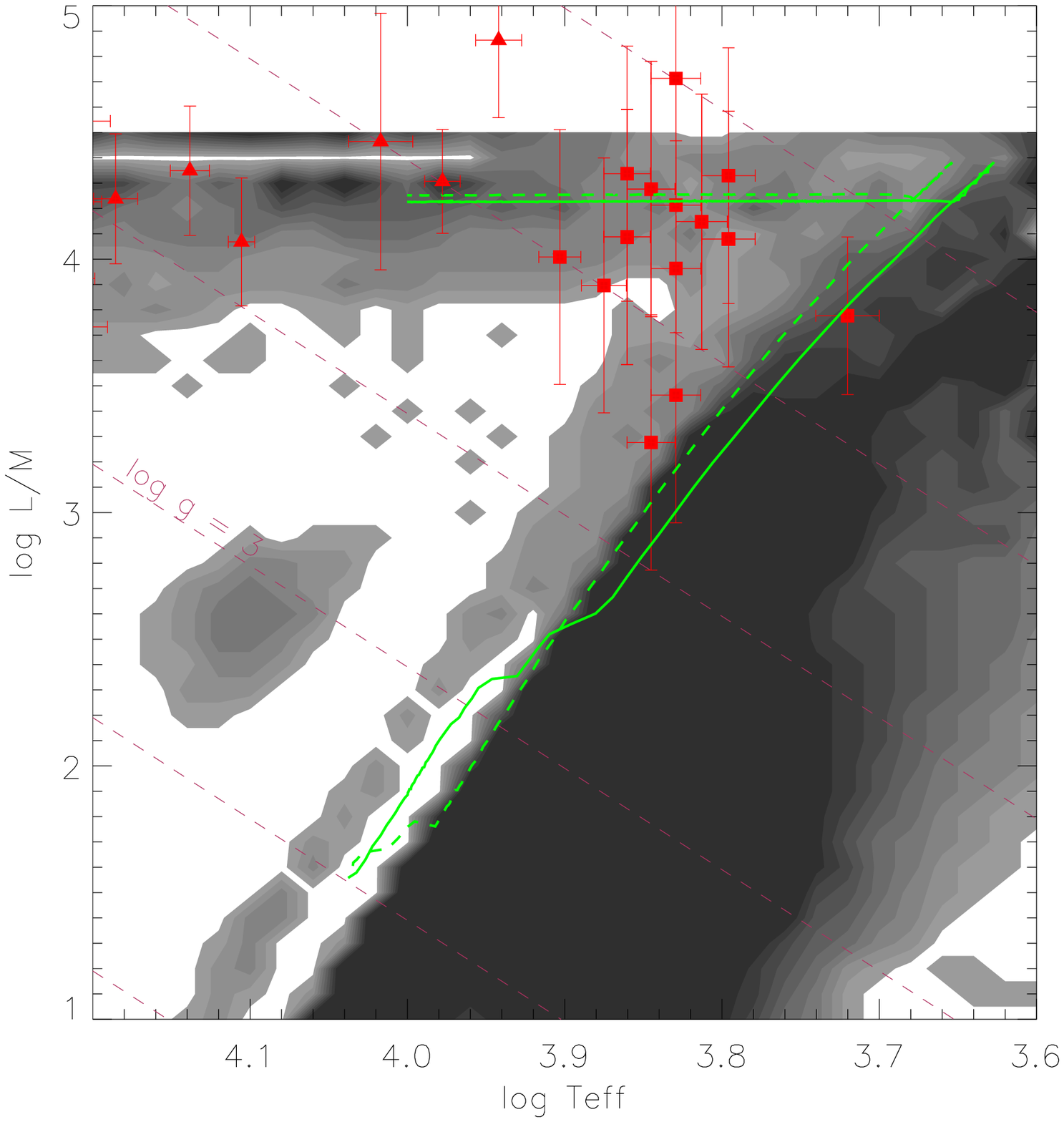}
 \caption{A grey-scale contour plot showing the number of unstable radial modes with $n<15$ in linear non-adiabatic pulsation
analyses of hydrogen-deficient stellar envelopes. The ordinate is $\log L/M$ in solar units, and the abscissa is $\log T_{\rm eff}$.
The plot is virtually invariant to the mass, at least in the range $0.2 - 1.0 {\rm M_{\odot}}$.
The solid squares  correspond to EHe stars, triangles to RCB stars. White means there are
no unstable modes, except for $\log L/M > 4.5$, where model envelopes are difficult to compute.
Black means 15 or more unstable modes. Surface
gravity contours for $\log g=5,4,3,\ldots$ are represented as broken lines. Evolution tracks for two
destroyed-disk post-merger models are superimposed (green): 0.55+0.30 \Msolar\ (solid) and 0.55+0.35 \Msolar\ (broken).
(Adapted from \citet{Jeffery2013b}).  }
   \label{f:puls}
\end{center}
\end{figure}

\section{Pulsation}

Following a series of studies of pulsation in hydrogen-deficient stars, \citet{Jeffery2013b}
extended their stability calculations as a function of hydrogen abundance (mass fraction: $X$) to a
larger range of effective temperature ($T_{\rm eff}$), luminosity-to-mass ratio ($L/M$) and $X$ than
before \citep{Jeffery99b}. Fig.~\ref{f:puls} shows a fragment of the instability domain for
$X=0.002$ and solar metallicity. At low \Teff\ the classical ``Cepheid'' instability strip is
evident, although shifted to higher \Teff\ because of the reduced surface hydrogen. At high $L/M$
strange modes are excited. The strange modes are less sensitive to $X$, and correspond to the
observed locations of the variable RCB and EHe stars. Strange-mode oscillations in stars with high
$L/M$ ratios have been identified for some time \citep{Wood76}. They appear theoretically in
non-adiabatic pulsation analyses and have no corresponding modes in the adiabatic approximation. In
particular, they appear to be associated with stellar envelopes where high opacities in the
ionisation zones also lead to a density inversion -- effectively creating a radiation-pressure
dominated cavity in the stellar interior. They are discussed at greater length by, {\it inter alia},
\citet{Gautschy90} and \citet{Saio98b}.

The evolution tracks for CO+He WD mergers lie entirely within the unstable domain, with pulsation in
classical radial modes driven by the second helium ionization zone at low \Teff, and in strange
modes at high $L$ (Fig.~\ref{f:puls}).

\section{Population Synthesis}

The birthrate of RCB stars from the merger of CO+He WD binaries has been estimated
using a binary population synthesis technique. We perform a Monte-Carlo simulation
to obtain a population of primordial binaries, then evolve each binary using a rapid binary evolution code (RBEC)
\citep{Hurley2000,Hurley2002} to obtain samples of double WD binaries. In the Monte Carlo
simulation, all stars are assumed to be members of binaries and have circular orbits, while the
primaries are generated according to the formula of \citet{Eggleton89}, {\it i.e.} following the initial
mass function of \citet{Miller79}  and in the mass range 0.08 to 100 \Msolar. The secondary mass, also with
a lower limit of 0.08\Msolar, is then obtained assuming a constant mass-ratio distribution. The
distribution of orbital separations, $an(a)$, is taken to be constant in $\log a$ for wide binaries
and falls off smoothly at small separations according to Eq. (12)
{\footnote { for $a \leq a_0$,  $an(a)=0.070(a/a_0)^{1.2}$;
for $a_0 \le a \le a_1$, $an(a)=0.070$, where $a_0=10 \Rsolar$, $a_1=5.75 \times 10^6 \Rsolar=0.13 \rm pc$. }} in \citet{Han98}, implying an equal
number of wide binary systems per logarithmic interval and approximately 50\% of stellar
systems with orbital periods less than 100 yr \citep{Han98}.

The numbers and characteristics of double WD binaries significantly depend on the critical mass
ratio for dynamical instability $q_{\rm c} (= m_{\rm 2}/m_{\rm 1})$ and common-envelope (CE) evolution
\citep{Han98}. Here we adopt equation (57){\footnote { for a normal giant, the
mass-radius relation $R \propto m^{-x}$  is assumed.
Thus $q_{\rm c}=[1.67-x+2(m_{\rm c1}/m_{\rm 1})]/2.13$, where $m_{\rm c1}$ is core mass of $m_{\rm 1}$.}}
of \citep{Hurley2002} for $q_{\rm c}$ when the mass donor is on
the first or asymptotic giant branch (AGB), and use the standard energy formalism
\citep{van76,Webbink84,Livio88} for CE evolution by setting $\lambda\alpha_{\rm ce} = 1.5$ which
reproduces the number of the double WD binaries in the Galaxy as in previous studies \citep{Wang2009}.

\begin{figure}
\centering \includegraphics [angle=-90,width=0.45\textwidth]{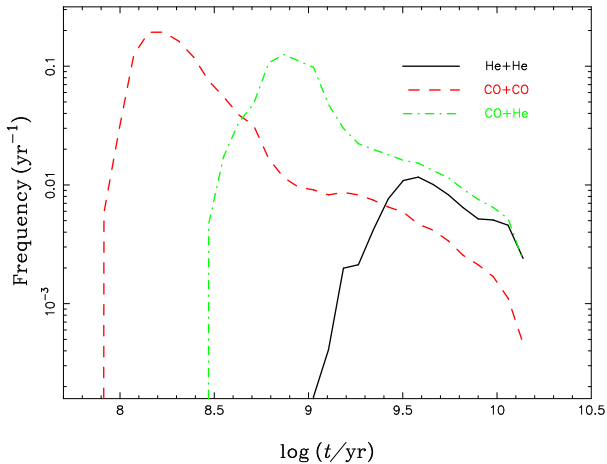}
\caption{Frequency of possible mergers from CO+CO, CO+He and He+He WD systems for a single
starburst (normalized to $10^{11}\Msolar$). The systems are assumed to have merged when the larger
components fill their Roche lobes. The three types of  white-dwarf merger are labelled.}
\label{cohe}
\end{figure}

\begin{figure}
\centering \includegraphics [angle=0,width=0.45\textwidth]{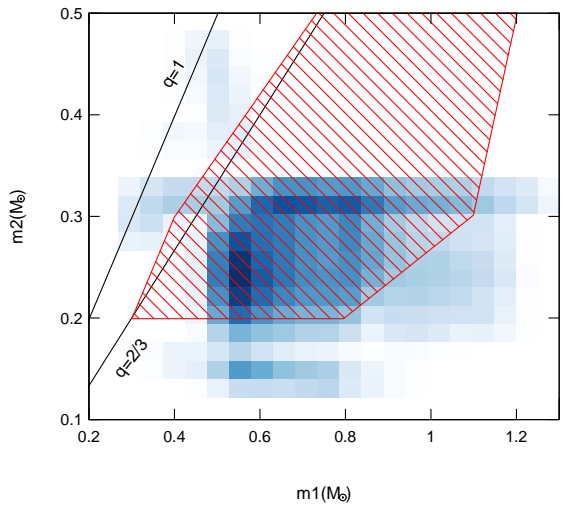}
\caption{ The relative number of CO+He binaries in
the $M_{\rm CO}-M_{\rm He}$ plane is illustrated in this colour-density plot,
where $m1$ represents CO WDs, and $m2$ represents He WDs. Otherwise the figure is similar to Fig.~\ref{binary}.
The hatched region marks the area studied by \citet{Dan2011}.
The  colour-density scale is linear, with the darkest shading corresponding to 3000 CO+He WD binaries per cell in the
$10^{11}\Msolar$ simulation. The sample comprises 345371 WD pairs.
}
\label{pop}
\end{figure}

Fig.~\ref{cohe} shows the rates of  possible mergers from CO+CO, CO+He and He+He WD binaries for
a single starburst (normalized to $10^{11}\Msolar$). These include only
mergers arising from WD binaries which fill their Roche lobes in a Hubble time.
If $q<q_{\rm c}$, mass transfer may be stable and may not produce a merger immediately.
The CO+He systems are produced from stable RLOF+CE\footnote{RLOF = Roche Lobe Overflow} or CE+CE
interactions to be sure that the orbital separation is small enough and the systems can merge in a Hubble time.
The He WDs {\it must} be formed in the second CE process.  The binding energy of an AGB-star envelope
(with a CO core) is small, so that CE ejection results in an orbital separation too wide to
merge in a Hubble time if the CO WD forms {\it after} the He WD.
Furthermore, the smaller the mass of the He WD, then the larger the binding
energy of the envelope and the more orbital energy is required and used for CE ejection.
Thus the CO+He binaries which have the shortest periods and hence the highest likelihood to merge
will be those containing the lowest mass He-WDs.
Consequently, most of the merged CO+He binaries in Fig.~\ref{cohe} have low-mass
He companions before merger.  Fig.~\ref{pop} shows a number-density plot for  CO+He WD binaries in the
$M_{\rm CO}-M_{\rm He}$ plane. If we adopt the critical mass ratio for dynamically unstable mass
transfer to be $q>0.6$ (\citet{Motl07} give 0.67), Fig.~\ref{pop} suggests that relatively few RCBs will be produced
from CO+He WD mergers.

\begin{figure}
\centering \includegraphics [angle=-90,width=0.45\textwidth]{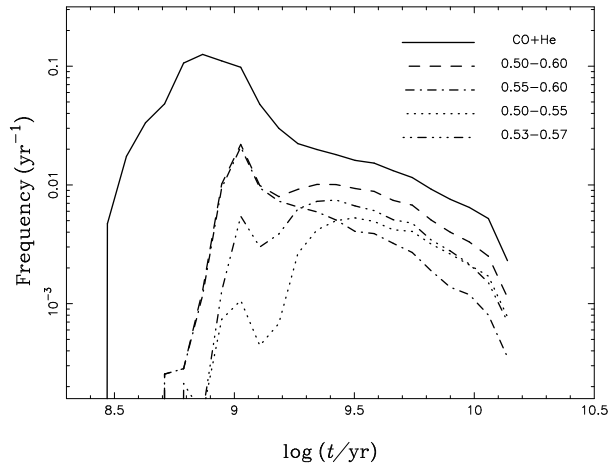}
\caption{Frequency of RCB stars from CO+He mergers for various ranges of the CO WD mass
and for a single starburst normalized to $10^{11}\Msolar$. CO+He refers to all possible CO+He mergers
as shown in Fig.~\ref{cohe}. }
\label{rate1}
\end{figure}

As discussed in \S2, mass transfer in many low mass-ratio binaries is probably unstable
and super-Eddington accretion may occur (Fig.~\ref{binary}: hatched region).
Indeed, \citet{Dan2011} find a weak coupling between spin and orbit that leads almost all systems with $q \gtrsim 0.2$
to merge. 
In our calculations, the \iso{12}{C} abundance of RCBs can only be explained if $M_{\rm CO}\approx 0.55\Msolar$.
In this case, mass transfer is unstable if $M_{\rm He}>0.2\Msolar$.
We have computed the RCB birthrate for various ranges of  the CO WD mass,
 i.e. for $M_{\rm CO}=0.5-0.6$, $0.55-0.6$, $0.5-0.55$ and
$0.53-0.57\Msolar$ (Fig.~\ref{rate1}). These show that RCBs appear at about $10^9$ yr after a single star burst,
and the peak frequency is $\approx 0.02\,{\rm yr}^{-1}$ for the optimum range of CO WD mass ($0.5-0.6\Msolar$).
The dip around 1.3 Gyr separates double WD binaries produced by the stable RLOF+CE process,
which dominate at long lifetimes, from the more compact double WD systems produced by the
CE+CE process, which dominate at short lifetimes. CO WDs with masses of $0.55-0.60\Msolar$
account for the formation of RCBs  when $t < 2$ Gyr,  while those with masses
of $0.50-0.55\Msolar$ dominate RCB production for longer lifetimes.

\begin{figure}
\centering \includegraphics [angle=-90,width=0.45\textwidth]{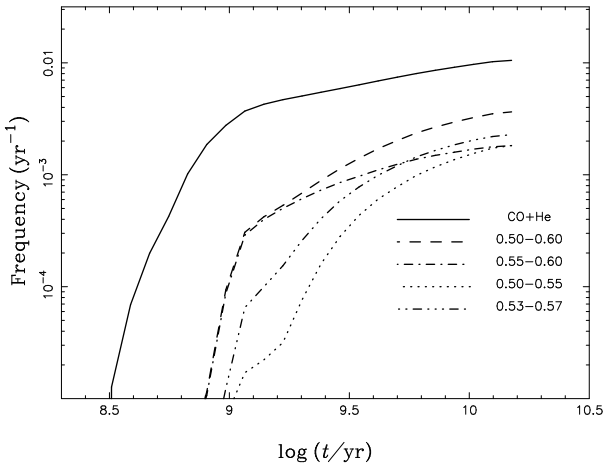}
\caption{As Fig.~\ref{rate1} for a constant star formation rate of 5 \Msolar/yr over the past 15 Gyr.  }
\label{rate2}
\end{figure}

Fig.~\ref{rate1} represents the formation rate from a single starburst, as possibly represented by
the old population of the galactic bulge to which RCBs and EHes are believed to belong.
The theoretical merger lifetimes are too short to explain the current numbers of RCBs and EHes in this way.
Sufficient numbers of CO+He WD binaries  can only come from relatively recent ($\leq 2$ Gyr)
star formation.  Fig.~\ref{rate2} shows the merger rates for a similar calculation assuming a constant
star-formation rate of $5\Msolar\,{\rm yr^{-1}}$ over the past 15 Gyr, resembling the Galactic disk.
The RCB birthrate in this simulation is $0.0036\,{\rm yr^{-1}}$ at 15 Gyr for the
optimum case  ($0.55-0.60\,\Msolar$),  enough to explain the number of RCBs in our Galaxy.

In our post-merger evolution models, the timescales of the RCB stages are $\approx$ 50\,000 to 70\,000 yrs.
Assuming a birthrate of $0.003\, {\rm yr^{-1}}$ gives $\approx 170 - 250$ RCB stars in the Galaxy.
As the timescale for the EHe  stage is less than 10\% of the RCB stage, the number of EHes will be $\approx 17 - 25$.
Current observations give 68 known active RCBs and 17 known EHes in the Galaxy. It is not clear what fraction of
stars in the theoretical RCB stage will be active in the sense of showing RCB-type minima, or what the overall
completeness statistics are.  The predicted and observed numbers are  comparable within these unknowns.

\section{Conclusion}

The merger of  a helium white dwarf with a carbon-oxygen white dwarf in a short-period binary
currently provides the preferred model for the origin of RCB stars.
Guided by hydrodynamical calculations of the dynamical merger, which provide approximate initial
conditions for the subsequent evolution, we have made one-dimensional calculations
of the post-merger evolution. We have considered the separate contributions of the hot
corona and cold Keplerian disk which are formed from the helium white dwarf,
and have also considered the chemistry of the helium white dwarf.

We have calculated how the luminosity and effective temperature of the star evolve over time, and
also how the surface composition behaves in response to the chemistry of the progenitor white dwarfs,
to nucleosynthesis during and after the merger, to disruption of the helium white dwarf and to convective
mixing during the post-merger evolution. We have included mass-loss  assuming a Bl\"ocker-type stellar
wind, and have considered the effects of enhancing this wind as a consequence of the hydrogen-deficient carbon-rich chemistry.
Following merger, the stars become luminous and spend some 50\,000 to 70\,000 yrs in the domain of the
RCB variables, where their surface chemistries strongly resemble those observed.
Our calculations suggest that the observed \iso{12}{C} abundances can only be produced by
a very small range in the mass of the CO white dwarf around $0.55\pm0.02 \Msolar$. As found previously,
\iso{18}{O} is produced during and survives the merger to account for observed overabundances of
oxygen. Some \iso{18}{O}, reacting with hydrogen from the helium white dwarf, and
\iso{3}{He} from the  same source, combining with \iso{4}{He}, can account for the enrichment of fluorine and lithium
seen in many RCB stars.

To account for the small-size of the Keplerian disk relative to the expanding star, and its likely ingestion into that envelope,
we have investigated a new class of model, the ``destroyed-disk" model, in which the disk is assumed to be totally destroyed
on a convective turnover timescale.  A CO WD mass of $0.55\pm0.02 \Msolar$ is still required to produce high surface carbon,  but
 a lower range in the He WD mass is permissible ($\geq 0.30\Msolar$).

The post-merger evolution tracks pass through a region of the $L-\Teff$ diagram where stars are predicted to be pulsationally unstable,
as is generally observed. Binary population synthesis (BPS) calculations  indicate that the highest fraction of CO+He WD binaries have masses
of around 0.55 (CO) and 0.25 (He) \Msolar\, respectively. Combining the evolution timescales with estimates of merger rates from the
BPS calculations yields an estimate of present-day numbers consistent with current observation.

A few problems still need to be addressed:
i)  post-merger evolution calculations and observed carbon abundances require the He WD masses to be in the range 0.30 \Msolar\ (destroyed-disk)
to 0.45 \Msolar\ (fast+slow) whereas BPS simulations show most He-WD companions have masses in the range 0.2 -- 0.3 \Msolar,
ii) observed luminosities and evolution calculations imply total masses around 0.8 -- 1.0 \Msolar,
whereas BPS simulations favour total masses in the range 0.75 -- 0.85 \Msolar\ (with some outliers in both cases),
  iii) the BPS calculations imply that spin-orbit coupling must be weak so as to allow dynamical mergers for
$0.5 < q < 0.67$.
and iv) the observed distribution of Galactic RCB  stars must be reconciled with a relatively young ($<2 $ Gyr) stellar population.
The first three might readily be solved by considering in detail the r\^ole of spin-orbit coupling
as suggested in \S\,9, and by \citet{Dan2011}. 
Accurate measurements of distances, luminosities and masses will provide crucial tests for the models,
whilst  work on  the effect of rotation and the distribution of angular momentum following
merger will have important consequences for interpreting the products.

A number of RCB observables have not been considered in the current paper.
The ejection of carbon-rich material from the system either during the merger or in the form of an enhanced Bl\"ocker-type
wind may lead to the formation of a cricumstellar shell. Cool dust shells are observed to exist around several
RCB stars. A comparison of the observed masses and sizes with what might be expected theoretically
would be valuable. It would be interesting to know whether the shells were formed during the stage A phase (as ejecta
from the dynamical merger) or during the subsequent stage B evolution.
More theoretical work on how mass-loss works in these stars would also
be useful, particularly in reconciling the {\it ad hoc} use of a  Bl\"ocker-type wind with the episodic dust ejection
associated with RCB variability.

In \S6 we  noted that CO+He WD mergers may also result in stars that are {\it not}
carbon-rich. One may speculate upon how such stars might appear and whether they could be recognised.
The evolution tracks continue to imply a passage through the AGB and post-AGB regions of the $L-\Teff$ plane.
With reduced carbon, the characteristic dust-ejection episodes and the  molecular
${\rm C}_2$ (Swan) absorption bands would be absent. The surface chemistry would remain
hydrogen-deficient, so CH and OH would not be strong, but other signatures such as lithium and fluorine
could persist. More detailed computations of the surface chemistry, based, for example,
on the calculations presented here, would be worthwhile.

\section*{Acknowledgments} The Armagh Observatory is supported by a grant from the
Northern Ireland Dept. of Culture Arts and Leisure.
Z.H. is partly supported by the Natural Science Foundation of China
(Grant Nos 11390374 and 11033008),
the Science and Technology Innovation Talent Programme of
the Yunnnan Province (Grant No. 2013HA005) and the Chinese Academy of
Sciences (Grant No. XDB09010202).
X.C. is partly supported by the Natural Science Foundation of China
 (Grant No. 11173055), the Talent Project of Young Researchers of Yunnan Province (2012HB037) and
the Chinese Academy of Sciences.

\bibliographystyle{mn} 
\bibliography{mybib} 

\label{lastpage}

\end{document}